\documentclass[12pt]{article}

\usepackage{cite}

\usepackage[nointlimits,reqno]{amsmath}
\usepackage{amssymb}
\numberwithin{equation}{section} 

\usepackage{graphicx}

\usepackage[pdftex,
 pdfproducer=TeXShop,
 pdfcreator=pdflatex]{hyperref}

\usepackage{xspace}
\usepackage{luty}

\usepackage{xcolor}

\begin{document}

\begin{titlepage}

\title{Phenomenology of Induced\\
Electroweak Symmetry Breaking}

\author{Spencer Chang}

\address{Department of Physics and Institute of Theoretical Science, University of Oregon\\ Eugene, Oregon 97403}

\author{Jamison Galloway}

\address{Center for Cosmology and Particle Physics, Dept. of Physics, New York University\\ New York, New York 10003}

\author{Markus A.~Luty, Ennio Salvioni, Yuhsin Tsai}

\address{Department of Physics, University of California, Davis\\
Davis, California 95616}

\begin{abstract}
\small
We study the phenomenology of models of electroweak symmetry breaking
where the Higgs potential is 
destabilized by a tadpole arising from the coupling to an
``auxiliary'' Higgs sector. 
The auxiliary Higgs sector can be either perturbative or 
strongly coupled, similar to technicolor models.
Since electroweak symmetry breaking is driven by a tadpole, the
cubic and quartic Higgs couplings can naturally be significantly
smaller than their values in the standard model.
The theoretical motivation for these models is that they can explain
the 125~GeV Higgs mass in supersymmetry without fine-tuning.
The auxiliary Higgs sector contains additional Higgs states that cannot
decouple from standard model particles, so these models predict a rich 
phenomenology of Higgs physics beyond the standard model.
In this paper we analyze a large number of direct and indirect constraints
on these models.
We present the current constraints after the 8~TeV run of the LHC, 
and give projections for the sensitivity of the upcoming 14~TeV run.
We find that the strongest constraints come from the direct searches
$A^0 \to Zh$, $A^0 \to t\bar{t}$,
with weaker constraints from Higgs coupling fits.
For strongly-coupled models, additional constraints come from 
$\rho^+ \to WZ$ where $\rho^+$ is a vector resonance.
Our overall conclusion is that a significant parameter space for such models
is currently open, allowing values of the Higgs cubic coupling down
to $0.4$ times the standard model value for weakly coupled models and vanishing cubic coupling for strongly coupled models.
The upcoming 14~TeV run of the LHC will stringently test this scenario and
we identify several new searches with discovery
potential for this class of models.
\end{abstract}

\end{titlepage}

\section{Introduction}
\label{sec:intro}
The experimental discovery of a Higgs boson by the 
ATLAS \cite{Aad:2012tfa} and CMS \cite{Chatrchyan:2012ufa} collaborations
at the Large Hadron Collider,
and the subsequent measurements of Higgs couplings
\cite{ATLAS-CONF-2014-009,Aad:2014eha,CMS:2014ega}
constitute revolutionary advances in particle physics.
In particular, the observed couplings of the
Higgs to $WW$ and $ZZ$ imply that the Higgs boson that has
been discovered is the dominant source of electroweak symmetry breaking (EWSB),
and may in fact be the sole degree of freedom in the Higgs sector.
Nonetheless, there are strong phenomenological and theoretical
motivations for studying the possibility of additional sources of 
EWSB.
The phenomenological motivation is obvious: it is essential to fully
test the standard model, the minimal model that can account for all
particle physics data, which predicts a single Higgs boson.
The theoretical motivation comes from the fact that models that
address the naturalness of the hierarchy between the electroweak
symmetry breaking scale 
and exponentially higher scales such as the
grand unification scale 
or Planck scale 
require extended Higgs sectors.
The most plausible possibilities are supersymmetry (SUSY) and
models where the Higgs is a pseudo-Nambu-Goldstone boson.
In both kinds of models, extended Higgs sectors are required as part
of their basic structure.

In this paper we study the phenomenology of \emph{induced electroweak
symmetry breaking} \cite{Azatov:2011ht,Azatov:2011ps,Galloway:2013dma}.
The defining property of this scenario is that the Higgs sector
is close to a limit where it reduces to two decoupled sectors.
We assume that only one of the sectors has Yukawa couplings
to standard model fermions, so Yukawa couplings
remain nonzero in this limit.
For the present introductory discussion, we consider the case where the
Higgs sector with Yukawa couplings consists of a single
weakly coupled scalar doublet, {\it i.e.}~the Higgs sector of the
standard model.
In the paper, we will consider supersymmetric models
with additional Higgs doublets, but the qualitative features are 
the same.
We refer to the Higgs sector without Yukawa couplings as the 
\emph{auxiliary Higgs sector.}

In the limit where the auxiliary Higgs sector decouples from the 
standard model Higgs, we assume that only the auxiliary Higgs sector
breaks electroweak symmetry.
That is, the standard model Higgs field has a positive quadratic
term.
When we turn on couplings between the two Higgs sectors,
these will in general induce a tadpole term for the standard model Higgs.
For example, in the simplest model where the
auxiliary Higgs sector consists of a single
doublet $\Si$, we have 
\beq
\label{eq:HSiMass}
V(H, \Si) = m_H^2 H^\dagger H - \bigl( \ep \Si^\dagger H + \hc\bigr) + \cdots,
\eeq
where $m_H^2 > 0$, and $\ep$ is a parameter with dimensions of mass squared
that couples the two Higgs sectors.
Provided we can neglect the higher order terms in minimizing the potential
with respect to $H$, we have
\beq
\avg{\Si} = \frac{1}{\sqrt{2}} \begin{pmatrix} 0 \\ f \end{pmatrix},
\qquad
\avg{H} = \frac{1}{\sqrt{2}} \begin{pmatrix} 0 \\ v_H \end{pmatrix},
\eeq
with
\beq
v_H = \frac{\ep}{m_H^2} f.
\eeq
To obtain the measured values of the $W$ and $Z$ masses, we must have
$v = \sqrt{v_H^2 + f^2} = 246\GeV$.
We see that in this class of models the VEV for $H$ is ``induced''
by its coupling to the auxiliary Higgs sector, which can even be larger than the inducing VEV if $\epsilon > m_H^2$.

The 125~GeV Higgs has properties close to the standard model Higgs.
To be consistent with ATLAS and CMS measurements of the $hWW$ and $hZZ$ couplings,
this requires $f \lsim 0.3v$ if the auxiliary Higgs sector is strongly coupled,
with somewhat larger values allowed in the weakly-coupled case.
On the other hand, the additional Higgs states in the auxiliary Higgs
sector must be sufficiently heavy not to be observed,
so we can write an effective theory where they are integrated out. 
In this effective theory, higher order terms in the coupling $\ep$ will be 
suppressed by powers of $\ep / m^2_\text{aux}$, where $m_\text{aux}$ 
is the mass of the heavy auxiliary Higgs states. 
Explicitly, we obtain for the light Higgs an effective potential of the form
\begin{equation}
V_\text{eff} = \frac{1}{2} m_H^2h^2 -\epsilon f h \left[
1 + c_1 \left(\frac{\epsilon}{m_\text{aux}^2}\frac{v_H}{f}\right)\frac{h}{v_H}
+ c_2 \left(\frac{\epsilon}{m_\text{aux}^2}\frac{v_H}{f}\right)^2 \frac{h^2}{v_H^2}+\cdots\right],
\end{equation}
where the terms of $O(h^3)$ and higher are suppressed provided that
\beq
\frac{\ep}{m_\text{aux}^2} \, \frac{v_H}{f} \ll 1.
\eeq
The coefficients $c_1$ and $c_2$ are expected to be of order 1;
for example, in the simplest 2 Higgs doublet model implementation, one finds
$c_1 = 1$, $c_2 = -\frac 12$.
Therefore the self-interactions of the light Higgs are naturally strongly 
suppressed, an important phenomenological feature of this class of models.
This motivates the study of parameter space where we can treat $\ep$
as a perturbation, and the VEV of $H$ can be viewed as being induced
by a tadpole.
Because the auxiliary Higgs sector has a small VEV and large physical
Higgs masses compared with the standard model Higgs sector,
the self-couplings in the auxiliary Higgs sector must be
stronger than the self-coupling of the standard model Higgs.

In addition to the phenomenological motivation,
induced EWSB is also
motivated by the problem of naturalness.
Supersymmetry (SUSY) gives an elegant and compelling solution to the
large hierarchy problem and predicts a light Higgs boson.
However, SUSY has a residual naturalness problem, namely that the Higgs
boson mass is generally predicted to be too light.
In the minimal supersymmetric standard model (MSSM), this
arises because the Higgs quartic is determined by the
electroweak gauge couplings to be $\la_H \sim g^2$, and the
lightest CP-even Higgs mass is given by $m_h^2 \sim \la_H v^2 \le m_Z^2$.
Loop corrections to $\la_H$ from top and stop loops
can raise the Higgs mass to the observed value,
but at the cost of $\sim 1\%$ tuning \cite{Okada:1990vk,Haber:1990aw,Ellis:1991zd,Brignole:1992uf,Carena:1995bx,Birkedal:2004zx}.
Some models that can generate a sufficiently large quartic
with improved naturalness include
non-decoupling $D$-terms \cite{Batra:2003nj,Maloney:2004rc}
and the next-to-minimal supersymmetric standard model
in special regions of parameter space \cite{Ellis:1988er,Espinosa:1992hp,Kane:1992kq,Hall:2011aa,Cao:2012fz}.
Induced EWSB offers a qualitatively different
solution to the naturalness problem, since the observed
Higgs mass gains a contribution from the original positive mass squared, 
rather than from an increased quartic.

There are several different possibilities for models of this kind.
One possibility is that the auxiliary Higgs sector is genuinely
strongly coupled, similar to a technicolor sector.
Technicolor models where strong interactions are the main source of 
the $W$ and $Z$ mass are definitively ruled out by the existence of a light Higgs.
Even before the Higgs discovery, such models suffered from severe phenomenological
problems, namely accounting for flavor mixing without flavor-changing neutral
currents, the large value of the top mass, and the absence of large
corrections to precision electroweak observables.
On the other hand, a technicolor-like auxiliary Higgs sector is motivated by the
Higgs discovery, and is free of the phenomenological problems of traditional
technicolor theories.
Complete supersymmetric models of this kind were constructed 
in \cite{Azatov:2011ht,Azatov:2011ps}.
The auxiliary Higgs has no couplings to fermions, so there are no
flavor problems associated with the strong dynamics.
The precision electroweak fit, relative to standard technicolor, is improved by the fact that the 
parameters that couple the strongly-coupled sector to the MSSM Higgs
bosons break custodial symmetry and generate a positive $T$ parameter
in addition to the (theoretically expected) positive $S$ parameter.
For minimal strong sectors, these corrections
are naturally within the experimentally allowed region.
The fact that strong EWSB occurs at the
SUSY breaking scale is naturally explained because
SUSY breaking forces the auxiliary Higgs sector away from a strongly
coupled conformal fixed point, so this is a UV-complete solution to the
SUSY naturalness problem.

Another possibility is that the auxiliary Higgs sector is perturbative,
although more strongly coupled than the electroweak gauge interactions.
Models of this kind were analyzed in \cite{Galloway:2013dma}.
In these models the large self-couplings in the auxiliary Higgs
sector can be generated either by $D$- or $F$-terms.
There is no conflict with precision electroweak measurements,
and the tuning in the EWSB is less
than $10\%$ in most of the phenomenologically allowed parameter
space.

We now turn from the motivation to the phenomenology of this
class of models.
In the limit where the auxiliary Higgs sector decouples,
the light Higgs degrees of freedom are the
longitudinal components of the $W$ and $Z$ coming from
the auxiliary Higgs sector, and the standard model Higgs doublet,
which has vanishing VEV in this limit, and therefore describes 4 physical scalars 
with a mass near 125~GeV.
When we turn on the coupling between the sectors, the fields
in the standard model doublet mix with the auxiliary Higgs fields,
but there are still 4 light scalar fields.
In addition to the CP-even 125~GeV Higgs state, there
is a neutral pseudoscalar $A^0$ and a charged scalar
$H^\pm$.
The new states from the auxiliary Higgs sector cannot be too
heavy because their mass is proportional to $f \lsim 0.3 v$,
and they cannot decouple because the mixing of these states with
the standard model Higgs is responsible for most of electroweak
symmetry breaking.
This class of models therefore has a very rich Higgs
phenomenology.

In this paper, we attempt to give a comprehensive study of 
the phenomenology of induced EWSB, for both strong and weakly
coupled auxiliary Higgs sectors.
One generic phenomenological prediction of this mechanism is that the
self-coupling of the Higgs is smaller than the standard model value.
Loop corrections to the Higgs quartic are large only when the theory is
fine-tuned, so a small Higgs quartic is directly motivated by naturalness.
For example, in minimal SUSY models the maximum value of the tree-level quartic 
is obtained for $\tan\be \to \infty$, and is about half of the standard model value.
A small quartic coupling implies a small cubic Higgs coupling, which reduces the destructive interference in Higgs pair production and 
thus can be observed
at the high-luminosity LHC \cite{Baur:2003gp,Baglio:2012np,Dolan:2012rv}.
On the other hand, as discussed above, this scenario also predicts additional
Higgs bosons with sizable couplings to standard-model particles,
and these are potentially observable with lower luminosity.

In order to have a well-defined parameter space for the searches,
we define phenomenological models to describe both strongly-coupled
and perturbative auxiliary Higgs sectors.
This allows us to compare the reach of different searches, and parameterizes
the coverage of these searches for this class of models in a physically meaningful
way.
To simplify the parameter space, we decouple one linear combination of the
MSSM Higgs fields $H_u$ and $H_d$ from the auxiliary Higgs sector.
In the first phenomenological model, the auxiliary Higgs sector consists of a nonlinear
realization of EWSB, with the addition of heavy vector
resonances near the scale $4\pi f$.
This is intended to model a strongly-coupled auxiliary Higgs
sector, as in \cite{Azatov:2011ht,Azatov:2011ps}. In the second model, 
the auxiliary Higgs sector is modeled by a single Higgs
doublet. This can be thought of as a limit of the weakly coupled models
discussed in \cite{Galloway:2013dma}.
After decoupling a linear combination of the MSSM Higgs fields,
this gives an effective 2-Higgs doublet model (type I) with a tractable
parameter space.

We analyze a large number of direct and indirect constraints on these models.
We include constraints coming from measurements and searches performed at the 8~TeV 
run of the LHC, and also make rough projections for the 14~TeV run.
We find that the strongest constraints come from
direct searches for $A^0 \to Z h$ and $A^0 \to t\bar{t}$.
Higgs coupling constraints are less constraining than direct searches,
and essentially the entire range of parameters probed by Higgs coupling measurements
is covered by direct searches.  

For weakly coupled models, we find that there is still a large parameter space allowed
by present constraints.
The 14~TeV LHC with 300~fb$^{-1}$ of integrated luminosity 
will probe a large amount of additional parameter space, but cannot completely
cover the full parameter space.
For strongly-coupled models, the parameter space is more fully covered.
This is mainly due to the fact that the branching ratio $A^0 \to Z h$ is
still significant even when $A^0 \to t\bar{t}$ is kinematically allowed,
so searches for $A^0 \to Z h$ are more constraining.
In strongly-coupled models there are also important constraints from heavy resonance 
decays such as $\rho \to WZ$.

We identify several searches that are presently not being done that
could have discovery reach in this class of models.
One is $A^0 \to t\bar{t}$ for $m_{t\bar{t}} < 500\GeV$.
This is a challenging search because a resonance near the $t\bar{t}$ threshold
has a complicated shape that must be carefully modeled.
Another is $\rho^+ \to W^+ A^0$ or $Z H^+$,
followed by $A^0 \to Zh$ or $t\bar{t}$, $H^+ \to t\bar{b}$.

One important benchmark for this class of models is the allowed suppression of
the Higgs cubic coupling $g_{hhh}$ compared to its standard model value.
This can be measured only with great difficulty at very high luminosity,
and one can ask whether this can be a discovery mode for this class of models,
or whether searches at lower luminosity will exclude or discover any model
with a large suppression.
Taking into account the 8~TeV data, we find that very large
deviations are still allowed, namely $g_{hhh} / g_{hhh}^\text{(SM)}
\gsim 0.4$ in models where the auxiliary Higgs sector is weakly-coupled,
and even smaller values for strongly-coupled models.
If there is no signal after 300~fb$^{-1}$ of 14~TeV running,
a deviation $g_{hhh} / g_{hhh}^\text{(SM)} \sim 0.7$ will still be
allowed in weakly-coupled models,
while in strongly-coupled models the entire range up
to $g_{hhh} / g_{hhh}^\text{(SM)} \sim 0.95$ will be covered
with only 20~fb$^{-1}$ by the $A^0 \to Z h$ search.

This paper is organized as follows.
In \S\ref{sec:simplifiedmodel}
we define the simplified models we use for our phenomenological study.
In \S\ref{sec:results} we present our results.
Our conclusions are summarized in \S\ref{sec:conclusions}.

\section{Simplified Models}
\label{sec:simplifiedmodel}
In this section we explain the simplified models that we use to study the
phenomenology of induced EWSB.
Although the emphasis in this paper is on the phenomenology and not the
model-building, we include some discussion of how these models are
related to complete supersymmetric models.

\subsection{Strong Induced Electroweak Symmetry Breaking}
We begin by discussing the models where the auxiliary Higgs sector is
strongly coupled and breaks electroweak symmetry at a scale $f$
\cite{Azatov:2011ht,Azatov:2011ps}.
To explain the coincidence of the strong coupling scale and the 
SUSY breaking scale, we assume that
the auxiliary Higgs sector is a strongly-coupled
conformally invariant theory.
SUSY breaking at the TeV scale then naturally
triggers confinement and EWSB
at the SUSY breaking scale.
SUSY is therefore not a good approximate symmetry in the strong
sector at the EWSB scale.

To avoid large corrections to the electroweak $T$ parameter, we assume
that the auxiliary Higgs sector respects an approximate custodial symmetry.
That is, the symmetry breaking pattern is
$SU(2)_L \times SU(2)_R \to SU(2)_{L + R}$,
with the electroweak gauge group embedded in the standard way.
We assume that the mass scale of strong resonances in this sector is
given by
\beq\eql{Lambdadef}
\La \sim \frac{4\pi f}{\sqrt{N}},
\eeq
where $N$ is a possible large-$N$ factor.
The precision electroweak corrections are proportional to $N$, motivating
$N \sim 1$.
Nonetheless, we keep $N$ as an adjustable parameter for generality.
We can also write \Eq{Lambdadef} as
\beq
m_\rho = g_\rho f,
\eeq
where
\beq
m_\rho \sim \La, 
\qquad
g_\rho \sim 4\pi/\sqrt{N}.
\eeq
In the effective theory below the scale $\La$, the only light modes from the
strong sector are the Nambu-Goldstone bosons, parameterized by a $2 \times 2$
unitary matrix $\Si$ transforming under $SU(2)_L \times SU(2)_R$ as
\beq
\Si \mapsto L \Si R^\dagger.
\eeq

The auxiliary Higgs sector is assumed to couple to the MSSM Higgs fields 
$H_u$ and $H_d$ via
\beq\label{eq:tadpolestrong}
\De\scr{L} = \la_u H_u \scr{O}_u^\dagger
+ \la_d H_d \scr{O}_d^\dagger,
\eeq
where $\scr{O}_{u,d}$ are operators from the strong sector
and $\la_{u,d}$ are couplings.
In complete SUSY models, these couplings can arise from
cubic superpotential couplings between the MSSM Higgs fields and 
composite operators quadratic in the ``quark'' fields in the strong sector.

The effective theory below the scale $\La$ was described in \cite{Azatov:2011ps}
for the case where both MSSM Higgs doublets are lighter than $\La$.
In this paper we consider a simplified limit where one linear combination
of $H_u$ and $H_d$ decouples, that is, has vanishing VEV and a mass of order 
$\La$ or larger.
In this case we can integrate out the heavy linear combination, and the effective 
theory below the scale $\La$ consists of a single light elementary Higgs doublet
\beq \label{mssmdec}
H = H_u \sin\be + \tilde{H}_d \cos\be
\eeq
($\tilde{H}_{d}=i\sigma^{2}H^{\ast}_{d}$) coupled to the Nambu-Goldstone modes 
from the strong sector.
Models with elementary Higgs doublets and technicolor have been studied since the 1990s
\cite{Samuel:1990dq},
but the focus was on the case where the Higgs masses were above the
electroweak breaking scale, and electroweak symmetry was dominantly broken by
technicolor dynamics.
(See however \cite{Carone:2006wj,Kagantalk}.)
Here we are focusing on the case where the dominant source of electroweak symmetry
breaking is the VEV of the light Higgs, and the role of the technicolor dynamics
is to induce a tadpole for the light Higgs.  
In the limit of exact custodial symmetry 
$\lambda_u \sin\beta = \lambda_d \cos\beta = \lambda$,
the leading terms in the low-energy effective theory are
\begin{align} 
\scr{L}_\text{eff} &= D^\mu H^\dagger D_\mu H - m_H^2 H^\dagger H 
- \la_H |H|^4+\cdots \nonumber \\
\label{lagrSC}
&+ \frac{f^2}{4} \tr (D^\mu \Si^\dagger D_\mu \Si)
+ c g_\rho f^3 
\left[
\la \tr(\Si^\dagger \scr{H}) + \hc  + O\bigl( (\la \scr{H}/m_\rho)^2 \bigr) \right]
+ \cdots
\end{align}
where 
$\la_H= \cos^2 2\beta (g^2+g'^2)/8$ and $\scr{H}$ is the $2 \times 2$ matrix
\beq
\scr{H} = \begin{pmatrix} \tilde{H} & H \end{pmatrix} \mapsto L \scr{H} R^\dagger.
\eeq
The Higgs fields can be parameterized by
\beq
H = \begin{pmatrix} a^+ \\
\frac{1}{\sqrt{2}} ( v_H + h + i a^0)  \end{pmatrix},
\qquad
\Si = e^{i\Pi/f}\,,\quad
\Pi = \begin{pmatrix} \pi^0 & i\sqrt{2}\, \pi^+ \\
-i\sqrt{2}\, \pi^- & -\pi^0 \end{pmatrix}.
\eeq
We normalize the coupling $\la$ so that the limit $\la \to g_\rho$ corresponds
to strong coupling at the scale $m_\rho$.
We then expect $c \sim 1$ in Eq.~(\ref{lagrSC}).
As discussed in the introduction, we assume  that the 
coupling $\la$ is small in the sense that $\la v_H / m_\rho \ll 1$.
In this case, we can neglect terms with higher powers of $H$ coupling
to $\Si$, as well as higher derivative terms in the effective
Lagrangian.

To gain some intuition for the dynamics, lets first consider the case of no quartic 
coupling for the light Higgs, which occurs for $\tan \beta=1$.
In this limit, the Higgs potential is the sum of a quadratic term
and a linear (tadpole) term, and
minimizing the Higgs potential gives
\beq
v_H = 2\sqrt{2} c \gap \frac{\la}{g_\rho} \gap \frac{m_\rho^2}{m_H^2} \gap f.
\eeq
The physical mass of the CP-even scalar is then
$m_h = m_H = 125\GeV$.
The coefficient of the linear term is determined by obtaining the correct
value for $v$, so the only undetermined parameter in the effective Lagrangian
in this approximation is $f$.  

If we allow the quartic coupling to be nonzero, we can solve for $m_H^2$ by extremizing in $h$:
\beq
m_H^2= \frac{2 \sqrt{2} c f^3 g_\rho \la- v_H^3 \la_H}{v_H}.
\eeq
Now the physical mass for the Higgs is 
\beq
m_h^2 
= m_H^2+ 3\la_H v_H^2 = \frac{2\sqrt{2} c f^3 g_\rho \la}{v_H} + 2\la_H v_H^2.
\eeq
The second term is the Higgs mass one finds by minimizing the standard model Higgs potential after replacing $v$ with $v_H$.  Thus, the coupling to the auxiliary sector has generated an additional correction to the mass 
\beq
\delta m_h^2 = \frac{2\sqrt{2} c f^3 g_\rho \la}{\sqrt{v^2-f^2}}
\eeq
where we have imposed that the correct amount of EWSB is generated by both sectors.  Taking into account this reduced amount of EWSB by the Higgs doublet leads to modified couplings of the Higgs to fermions and electroweak gauge bosons
\beq
\label{eq:hcoupstrong}
\ka_f = \frac{g_{h \bar{f} f}}{g_{h \bar{f} f}^\text{(SM)}} 
= \frac{1}{\sqrt{1-f^2/v^2}},
\qquad 
\ka_V = \frac{g_{hVV}}{g_{hVV}^\text{(SM)}} = \sqrt{1-f^2/v^2},
\eeq
where $V = W, Z$.
Due to the genuinely different shape of the potential, the Higgs cubic coupling is strongly modified compared to the SM
\beq
\ka_h = \frac{g_{hhh}}{g_{hhh}^\text{(SM)}}  = \frac{\la_H}{\la_\mathrm{SM}}\sqrt{1-f^2/v^2}
\eeq
where $\la_\mathrm{SM}=m_h^2/2v^2$.
As expected, $\ka_h = 0$ in the limit where the $H$ quartic vanishes.

The effective theory also contains a triplet of pseudoscalars that are a
linear combination of the CP-odd modes in $H$ and the Nambu-Goldstone
modes in $\Si$.  
The mass matrices for the neutral $(a^0, \pi^0)$ and charged $(a^\pm, \pi^\pm)$
scalars are equal in this approximation.  In addition, in the limit where we decouple the two sectors by taking $\la\to 0$ we should find two sets of Goldstone bosons, 
which explains why the mass matrices end up proportional to $\delta m_h^2$:
\beq
\scr{M}^2 = \delta m_h^2 \begin{pmatrix} 1 & v_H/f \\ v_H/f & v_H^2 / f^2 \end{pmatrix}.
\eeq
This has a zero eigenvector corresponding to the linear combination
that is eaten by the $W$ and $Z$, and
the physical combinations  orthogonal to the Goldstones
\beq
A^0 = \frac{1}{v} \left( f a^0 + v_H \pi^0 \right),  \quad H^+ = \frac{1}{v} \left( f a^+ + v_H \pi^+ \right)
\eeq
which have degenerate masses
\beq \label{mASC}
m_{A}^2 =m_{H^+}^2 = \delta m_h^2 \frac{v^2}{f^2}. 
\eeq
Since $\delta m_h^2 \lsim (125 \GeV)^2$, this gives an upper bound of $m_A \lsim 125\, (v/f) \GeV$.
For $f < v_H$, the physical pseudoscalars are dominantly composite states, but still have reduced couplings to the CP even Higgs, gauge bosons,
and fermions determined in terms of $f$:
\begin{align}
g_{A^0 h Z} &= \frac{g}{2\cos\theta_W} \frac{f}{v},
\\
g_{A^0 \bar{f} f} &= \pm \left(\frac{m_f}{v_H}\right)\left(\frac{f}{v}\right) i \ga_5,
\label{SCAcoup}
\\
g_{H^- t \bar{b}} &= \sqrt{2}\left(\frac{f}{v}\right) 
\left(\frac{m_t}{v_H} P_L- \frac{m_b}{v_H} P_R \right).
\label{SCHpcoup}
\end{align}
which have a structure similar to a Type-I two Higgs doublet model.  

\begin{figure}[t!]
\centering
   \includegraphics[width=.47\textwidth]{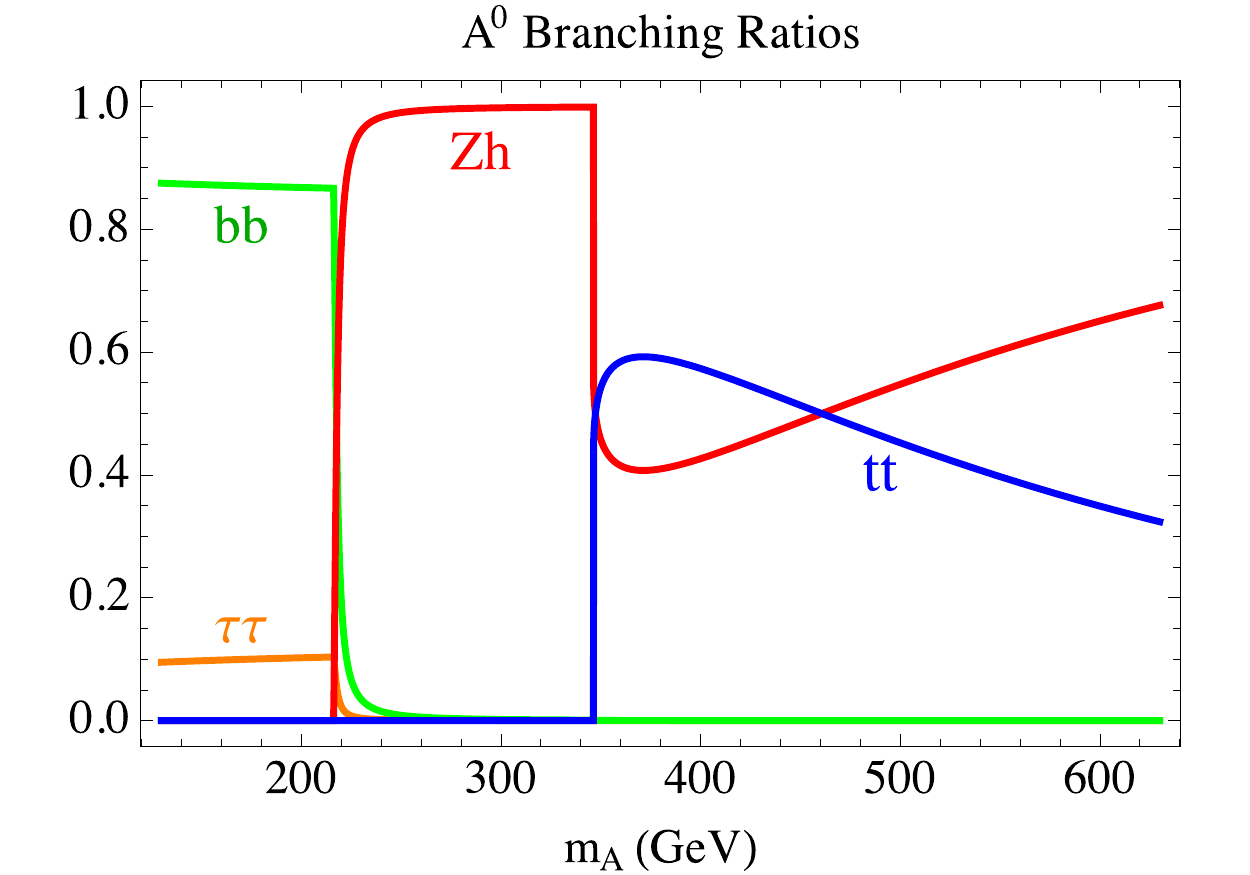} 
\quad 
   \includegraphics[width=.47\textwidth]{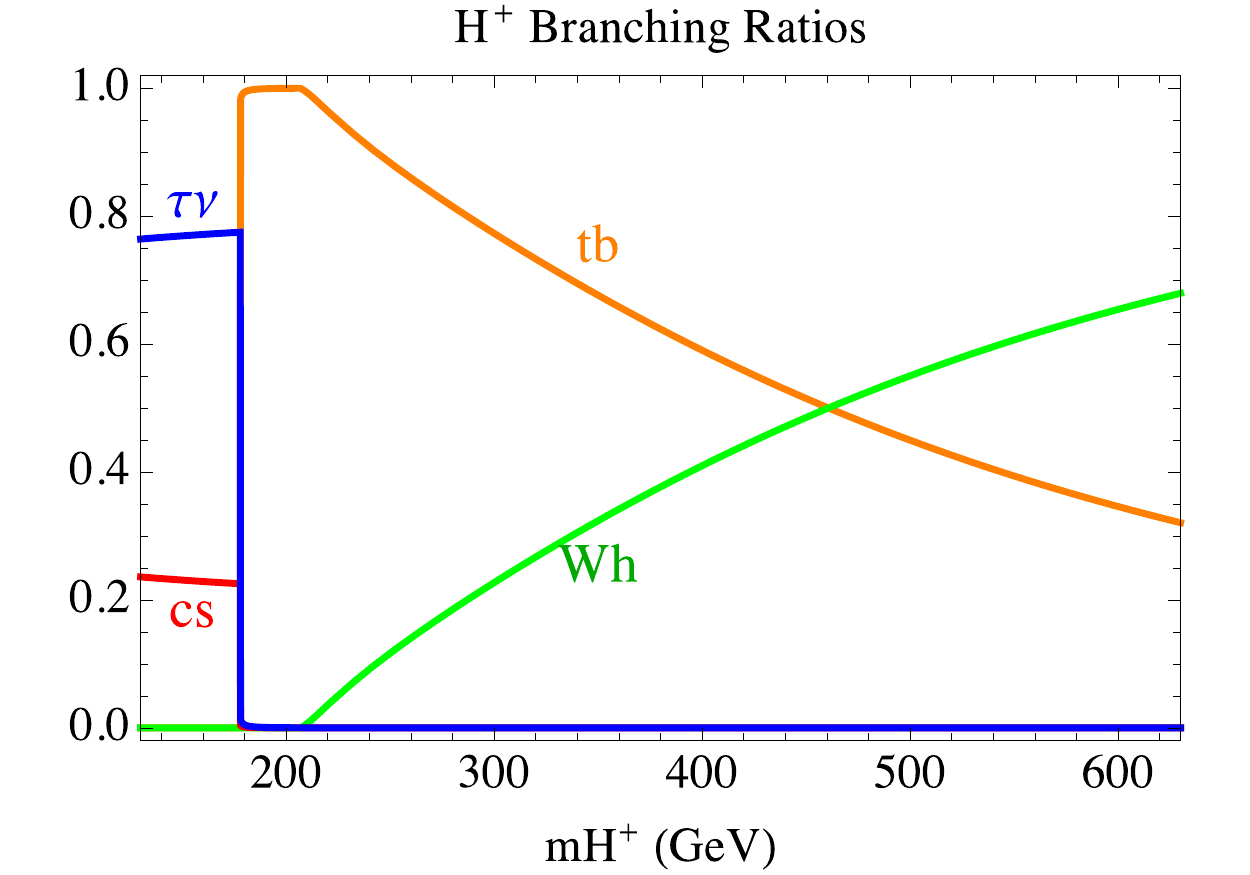}
\caption{\small{Branching ratios for $A^0$ and $H^+$ for $\la_H=0$.  The results are not strongly dependent on the actual value of $\la_H$.}}
\label{fig:strongpseudobrs}
\end{figure}

One can trade the Lagrangian parameters for the more physical parameters of $m_A, v, m_h, \la_H$.  Fixing the electroweak VEV and Higgs mass to their observed values, we are left with $\la_H$ and $m_A$.   In terms of these parameters, the amount of breaking in the strong sector is  
\beq
f= v\, \sqrt{\frac{1-\la_H/\la_\text{SM}}{m_A^2/m_h^2- \la_H/\la_\text{SM}}},
\eeq
which goes to $v$ as $m_A \to m_h$ and shows that we should consider the range $\la_H/\la_\text{SM} \in [0,1]$.
In Fig.~\ref{fig:strongpseudobrs}, we show the branching ratios of the pseudoscalar states when $\la_H=0$.  The values of the branching ratios are only weakly dependent on $\la_H$ and thus the phenomenology of these states is mainly dependent upon their mass.  
An important feature of the strongly-coupled scenario is that the $A^0\to Zh$ 
branching ratio remains large even as one goes above the top quark threshold.  
This does not occur for the weakly coupled model (see Fig.~\ref{fig:weakpseudobrs}), 
and explains why the $A^0 \to Zh$ search is more constraining in the strongly-coupled case.

In addition to the light fields of the theory,
the LHC can probe the heavy resonances of the strong
sector.
Higgs coupling fits require $f \lsim 0.3 v$ (see below)
so the mass of these resonances is expected to be near or below the TeV scale.
We explore this phenomenology with the phenomenological model of vector
resonances of \cite{Falkowski:2011ua},
which is based on the approach of \cite{Bando:1984ej,Bando:1987br,Casalbuoni:1985kq}.  

The effective Lagrangian has two dimensionless free parameters, 
$g_\rho$ and $\alpha$.  
In terms of these parameters, the $\rho$ has the following properties
\beq
m_\rho = g_\rho \sqrt{\alpha} f, \qquad g_{\rho \pi \pi} = g_\rho \alpha/2.  
\eeq
For the QCD $\rho$, these values are $g_\rho \simeq 6.4$, $\alpha \simeq 1.7$.
In general we expect $g_\rho \sim 4\pi/\sqrt{N}$ and $\al \sim 1$, and we will allow 
these parameters to vary in the phenomenology below.  
Integrating out the $\rho$ gives a contribution to the $S$ parameter
\beq
\De S_\rho \simeq 0.2 \left( \frac{7.9}{g_\rho} \right)^2.
\eeq
Precision electroweak measurements give $S < 0.2$ at $95\%$ confidence level,
so we see that the QCD $\rho$ is marginal.\footnote{The measurement of trilinear gauge couplings at LEP2 \cite{Schael:2013ita} gives a weak constraint on $g_\rho$, $g_\rho\gsim 1.5$ at $95\%$ CL. 
}
Taking the largest value of $f$ allowed by Higgs couplings, $f\sim 70\;\mathrm{GeV}$, we expect $500\GeV \lsim m_\rho \lsim 900\GeV$, where the lower bound comes from precision electroweak constraints and the upper bound from perturbativity.

\subsection{Weakly-Coupled Induced Electroweak Symmetry Breaking}
We now turn to models where the auxiliary sector is perturbative.  
The important new feature here is
the presence of neutral CP-even modes originating from the auxiliary states which are absent in the
strongly-coupled model. With weakly coupled $\Si$ fields, fluctuations about $f$ are physical and will
partially comprise the light Higgs, introducing a single additional mixing parameter and accordingly 
affecting the phenomenology of the scalars in the IR.  

Here we consider a simplified limit where the low energy theory consists of two doublets:
$H$ of Eq.~\eqref{mssmdec}, and a single auxiliary state $\Si$.  
The effective potential for these doublets is as in 
Eq.~(\ref{eq:HSiMass}), now including all terms relevant for obtaining the vacuum state:
\beq\label{WCpotential}
V_\text{eff} 
= m_H^2 H^\dagger H+m_\Si^2 \Si^\dagger \Si - (\ep \Si^\dagger H  + \hc)
+ \la_{\Si}|\Si|^4  + V_{D}
\eeq
Here $V_{D}$ denotes contributions from the $SU(2)_L\times U(1)_{Y}$ $D$-terms, and
the mass mixing is traced to couplings of $H_{u,d}$ to $\Si$ via $\ep = \ep_{u}\sin\be + \ep_{d}\cos\be$ with
the angle $\beta$ defined by $\tan\beta = v_u/v_d\,$.
Note that $V_D$ is set by $\tan \beta$, so fixing the masses of the light Higgs and the weak gauge bosons
leaves just two free parameters in this theory.
The additional auxiliary self-interaction $\la_\Si$  can arise from non-minimal $F$- or $D$-terms, 
as considered in various UV-complete models \cite{Galloway:2013dma}.  In the present
case, we will be concerned only with the fact that $\la_\Si$ can be substantially larger than
the SM $D$-term contributions,  allowing a sensible tadpole-like limit for the EFT.

The $D$-terms of the SM  group have relevant phenomenological implications and will be 
consistently included in our analysis. 
Most importantly, they can give a sizable contribution to the cubic coupling of the light 
Higgs, depending on the size of $\tan\be$. 
Because we are not relying on a large quartic for $H$, there is no preference for
large $\tan\be$ from naturalness arguments.
We therefore present results for two representative cases: 
$\tan\be = 1$, which minimizes the $D$-term contribution to the potential of $H$, 
and $\tan\be = \infty$, which maximizes it. 

\paragraph{$\boldsymbol {\tan  \beta = 1}$:}  
We consider first the limit $\tan \beta =1$, where the light
$H$ doublet lies along a $D$-flat direction.  
The SM $D$-terms thus generate only $V_D = \la_Z |\Si|^4$, where we define
$$
\la_Z = \frac{g^2 + g'^{\, 2}}{8}\,.
$$
This case therefore realizes dominance of the auxiliary self couplings in the most obvious way, and provides the clearest illustration of the perturbative model's parametrics.

First, as in the strongly-coupled model, there is a triplet of pseudoscalars
with mass
\beq
\label{eq:tb1triplet}
m_A^2 =  m^2_{H^\pm} = \frac{v^{2}}{f^{2}} m_{h}^{2} \left(1 + \frac{m_{h}^{2}v_H^2}{2(\la_{\Si}+\la_{Z})f^4 - m_h^2v^2}\right) .
\eeq
There is additionally the heavy CP-even neutral scalar, which is characteristic of the weakly coupled case as described above.  Its mass is given by
\beq
\label{eq:tb1H}
m_{H^0}^2 = 2(\la_{\Si}+\la_{Z})f^2 + m_A^2 -m_h^2.
\eeq 
For a given $m_A$, the ratio $f/v$ is thus completely determined by obtaining the correct mass for the light Higgs state.  

From the limit $m_A \to \infty$, $m_{H^0} \to \infty$ in Eqs.~(\ref{eq:tb1triplet})
and (\ref{eq:tb1H})
we observe the decoupling limit of the model, where 
$\la_{\Si} \to m_h^2 v^2/2 f^4-\la_Z$ from above.  
Note that Eq.~(\ref{eq:tb1H}) implies $m_{H^0}> m_A$ in the full parameter space,
where the splitting of these states becomes large as we take $\la_\Si>\la_Z$.
This is an important distinction with respect to the EWSB sector of the MSSM, 
where $m_{H^0}^2 \leq m_{A}^2 + m_Z^2$ at tree-level.

The light Higgs here contains an admixture of auxiliary Higgs sector states,
modifying its couplings.
Its coupling ratios are given by 
\beq \label{hcouptb1}
\ka_{f} \simeq 1 + \frac{m_h^2}{m_A^2}\,,\qquad \ka_{V} \simeq 1 - \frac{m_h^4}{2m_A^4}\left(\frac{\sqrt{2(\la_\Si+\la_Z)}\,v}{m_h}-1\right)\,,
\eeq
where we write the leading terms in the expansion for small $m_h^2/m_A^2$. 
(This  expansion is more reliable than the expansion in $f/v$ because larger values of 
$f$ are  allowed in the weakly coupled models.)
The coupling to fermions receives the larger correction 
and thus drives the experimental constraints. 

We find that the cubic coupling of the Higgs is subject to the largest fractional deviations from the SM.
Parameterizing the ratio $m_h^2 / m_A^2$ by use of $\ka_f $ in Eq.~(\ref{hcouptb1}),
we find for $\la_\Si \lsim 2$ a rescaling
\beq \label{cubicWC}
\ka_h - 1 \simeq -2\left(\frac{\sqrt{2(\la_\Si + \la_Z)}\,v}{m_h}-1\right)
(\ka_f - 1).
\eeq
This shows that for $\la_\Si \sim 1$ the Higgs self-coupling receives a parametrically 
larger correction than the fermionic coupling. 
For example, for $\la_\Si \simeq 2$ we obtain $\ka_h - 1 \simeq -6.4 (\ka_f - 1)$.
This allows a very non-standard cubic coupling even when the vector and Yukawa coupling fit 
constraints are satisfied. 
Higher order terms in Eq.~(\ref{cubicWC}) become important for larger $\la_\Si$, but substantial deviations from
$\ka_h = 1$ persist.

\begin{figure}[t!]
\centering
   \includegraphics[width=.47\textwidth]{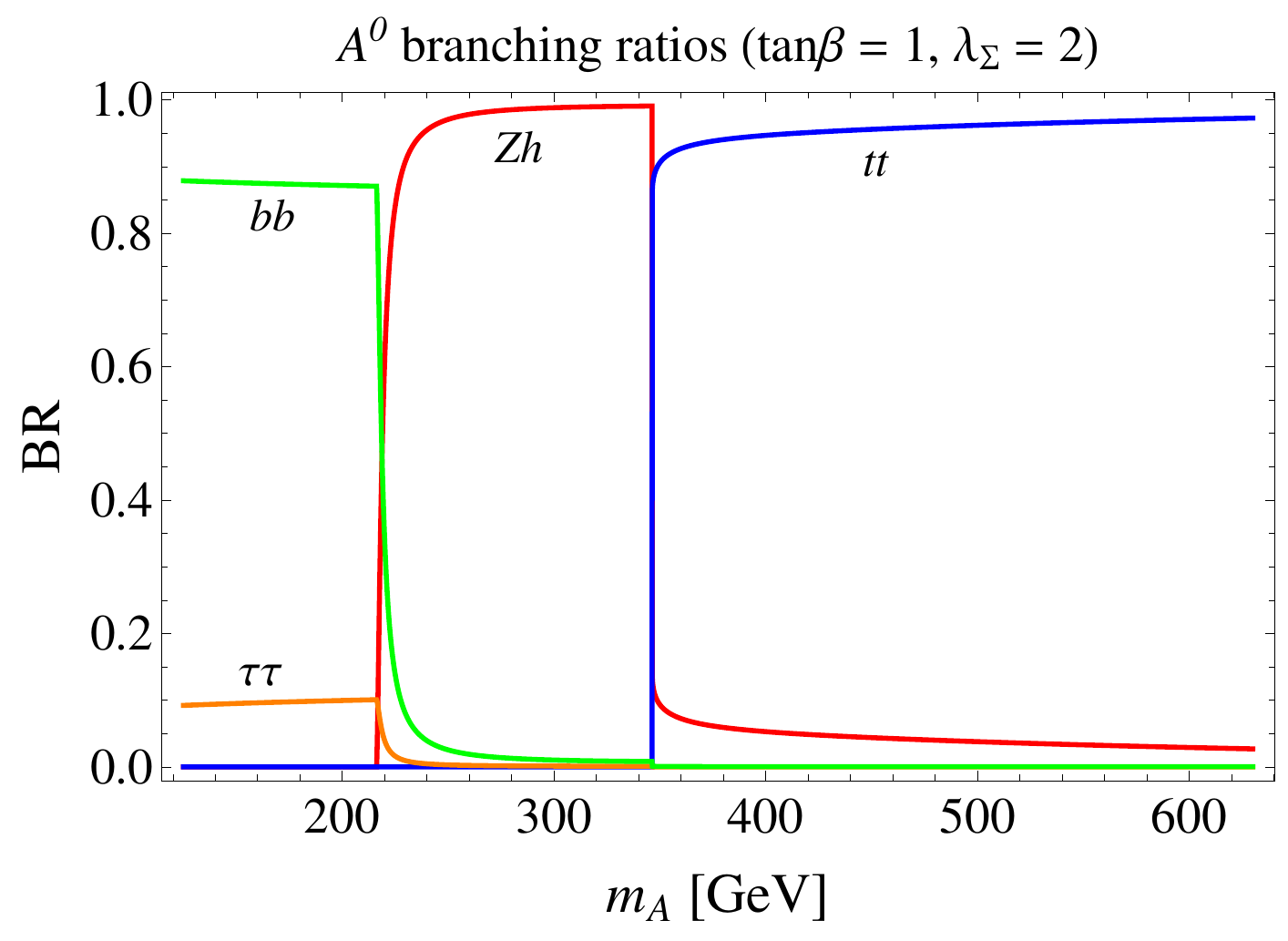} 
\quad 
   \includegraphics[width=.47\textwidth]{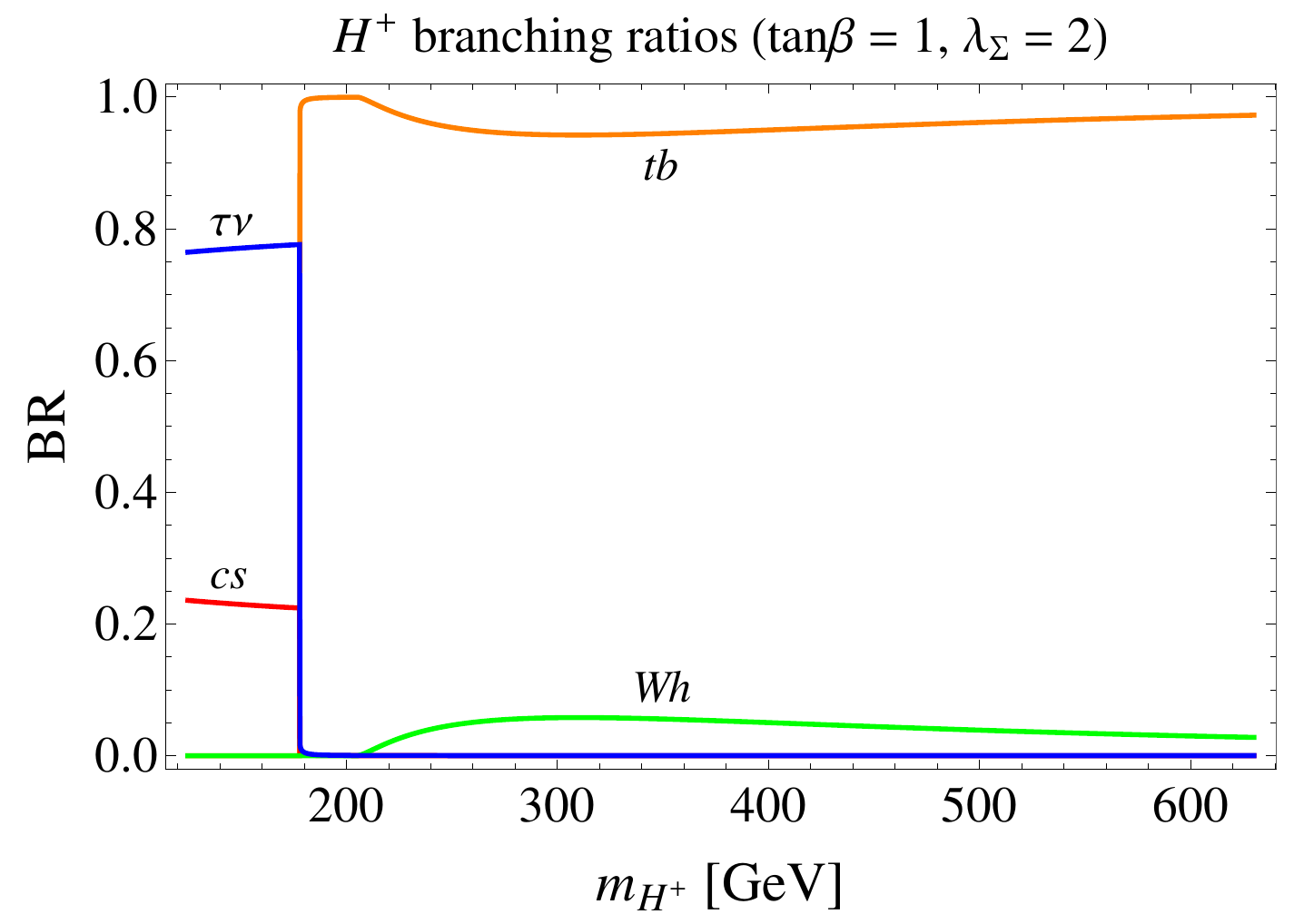}
\caption{\small{Branching ratios for $A^0$ and $H^+$ in the weakly coupled model with $\tan\beta =1$. The auxiliary quartic is fixed to $\la_\Si=2$. The results are not strongly dependent on the actual value of $\la_\Si$ within the perturbative region.}}
\label{fig:weakpseudobrs}
\end{figure}

In the induced tadpole region, $f$ is sizable and therefore the pseudoscalar triplet is 
rather light, making the direct searches of $A^0$ into $\tau\tau, Zh, t\bar{t}$ the 
dominant direct constraints on the model. 
The couplings of the pseudoscalars to fermions have the same expression as in 
Eqs.~(\ref{SCAcoup}, \ref{SCHpcoup}), while 
\beq
g_{AZh} = \frac{g(f \cos\ga - v_H \sin\ga)}{2 v \cos\theta_W},
\eeq
where $\ga$ is the mixing angle between the CP-even states. The branching ratios of the neutral $A^0$ and charged $H^+$ are shown in Fig.~\ref{fig:weakpseudobrs}, 
where the dominance of the decays $A^0\to t\bar{t}, H^+ \to t\bar{b}$ at large $m_A$ is evident. 
This feature is present for any perturbative $\la_\Si$ and is in contrast with the strongly 
coupled model, where $A\to Zh, H^+ \to Wh$ are largest (see Fig.~\ref{fig:strongpseudobrs}). 
There is no inconsistency in these results: it is easy to verify that in the limit 
$\la_\Si \to\infty$ the weakly coupled model with $\tan\beta =1$ reproduces exactly the 
strongly-coupled model in Eq.~\eqref{lagrSC} with vanishing Higgs quartic, $\la_H = 0$.

\begin{figure}[t!]
\centering
   \includegraphics[width=.47\textwidth]{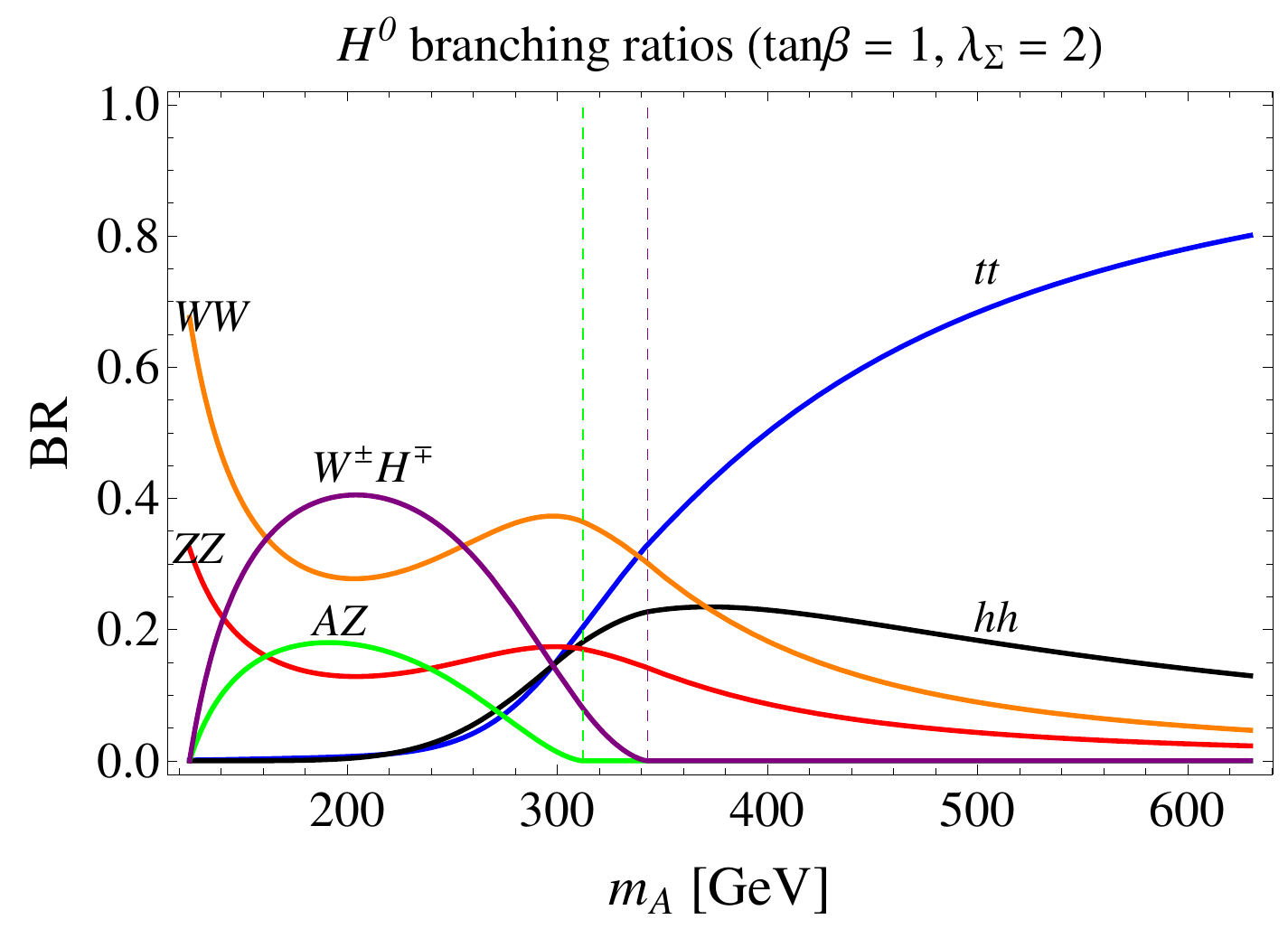} 
\quad 
   \includegraphics[width=.47\textwidth]{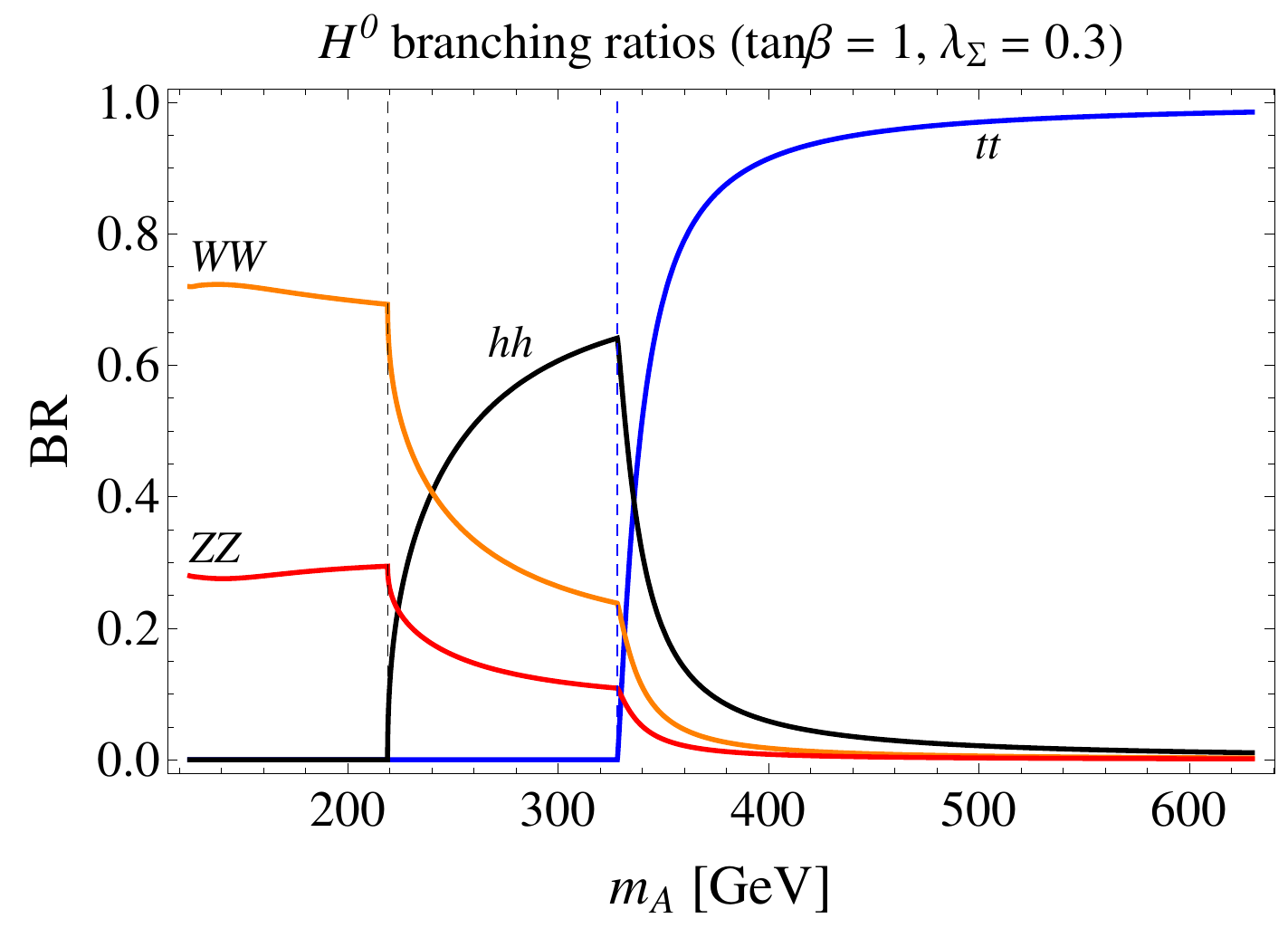}
\caption{\small{Branching ratios for $H^0$ in the weakly coupled model with $\tan\beta =1$, 
for $\la_\Si=2$ (left) and $\lambda_\Sigma = 0.3$ (right). 
The vertical dashed lines indicate the thresholds where new decay channels open up.}}
\label{fig:weakH0brs}
\end{figure}

The $H^0$ is typically much heavier than the triplet and has a relatively 
small production rate at colliders. 
Its decays depend more sensitively on $\la_\Si$ and are shown in Fig.~\ref{fig:weakH0brs} for two representative cases, one with larger quartic where the decays $H^0 \to A^0 Z, H^\pm W^\mp$ are open, and one with smaller quartic where these decays are kinematically inaccessible. 

\paragraph{$\boldsymbol {\tan  \beta = \infty}$:} 
For $\tan\be=\infty$ the $D$-terms are 
\beq
V_{D} \big|_{(t_\be=\infty)}=\la_{Z}(H^\dagger H-\Si^\dagger \Si)^2 + \frac{g^2}{2} H^\dagger \tilde{\Si} {\tilde{\Si}}^\dagger H\,.
\eeq
The main consequence is that the Higgs cubic coupling is significantly larger than in the 
$\tan\be=1$ case. 
In addition, the masses of the charged Higgs and pseudoscalar are split as 
$m_{H^\pm} = \sqrt{ m_{A}^2 + m_W^2 }$, which slightly relaxes the bounds on 
$H^{\pm}$ such as $b\to s\ga$, $R_b$ and $t\to H^+ b$. 
The Higgs couplings are also modified: the coupling to fermions is 
\beq \label{hcouptbinfty}
\ka_{f} \simeq 1 + \frac{m_h^2}{m_A^2}\left[1+\frac{m_Z^2}{m_h^2}\left(\sqrt{\frac{2}{\la_\Si}}\frac{m_h}{v}-1\right) + O(\la_Z^2)\right]\,,
\eeq
while $\ka_V$ again deviates from the SM only at $O(m_h^4/m_A^4)$.

\section{Results}
\label{sec:results}
In this section we discuss the current experimental constraints on the models, as well as the projected sensitivity of the $14$ TeV LHC in testing their parameter space.

\subsection{Strong Induced Electroweak Symmetry Breaking}
\begin{figure}[t!]
\centering
\includegraphics[width=.6\textwidth]{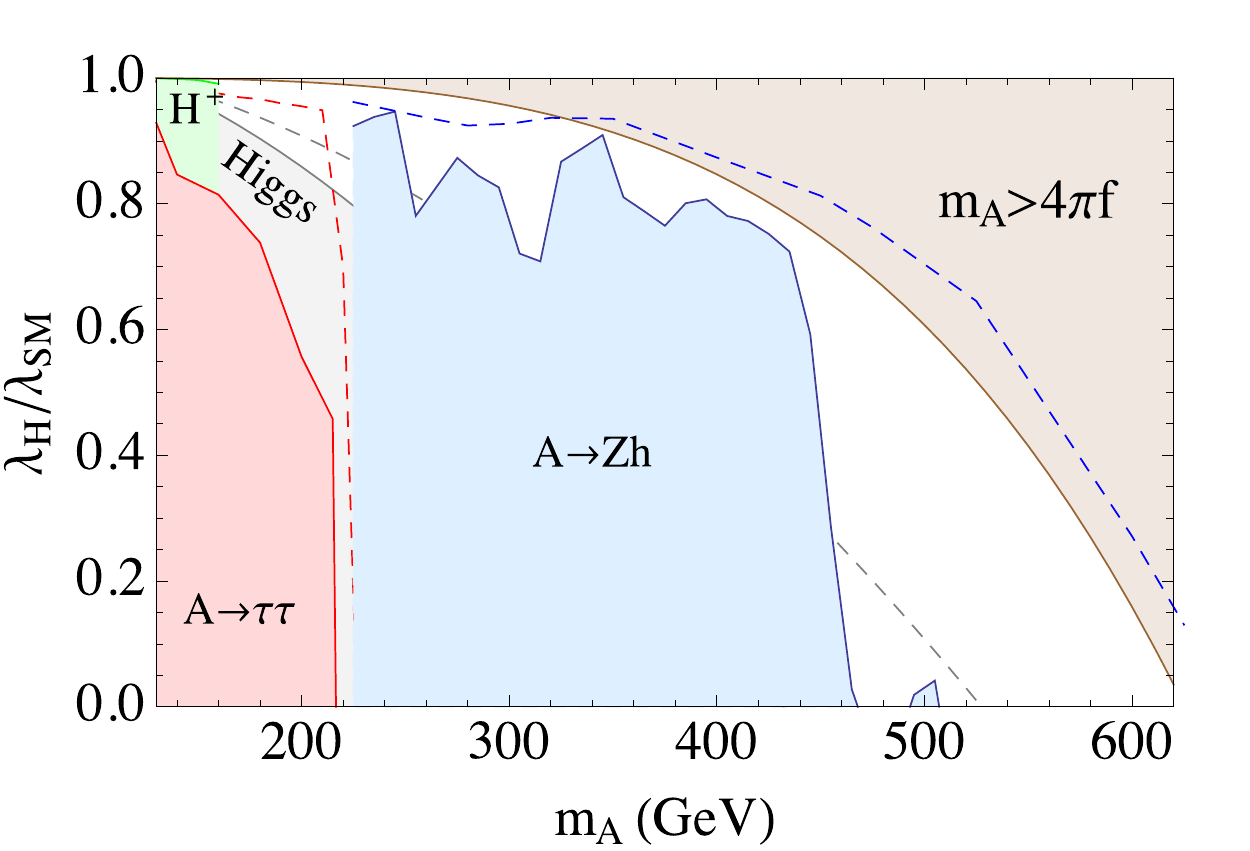} 
\caption{\small{Direct constraints on the $A^0$ and $H^+$ for strong induced EWSB.  The light solid gray is the limit from the combined LHC Higgs coupling fit.  The solid shaded regions represent limits from LHC searches for $A\to Zh, A\to \tau\tau, t\to H^+b\to (\tau^+ \bar{\nu}) b$.  The dashed lines show projections for the Higgs coupling constraint, $\tau\tau$ and $Zh$ search at the 14 TeV LHC, assuming respectively 300 fb$^{-1}$ for the coupling fit and $\tau\tau$ search and 20 fb$^{-1}$ for $Zh$.  Finally, the shaded region in the upper right is where the effective theory breaks down due to the particles being above the strong coupling scale of the nonlinear sigma model for $\Si$. }}
\label{fig:strongpseudolimits}
\end{figure}

\begin{figure}[t]
\centering
\vspace{-3mm}
   \includegraphics[width=0.6\textwidth]{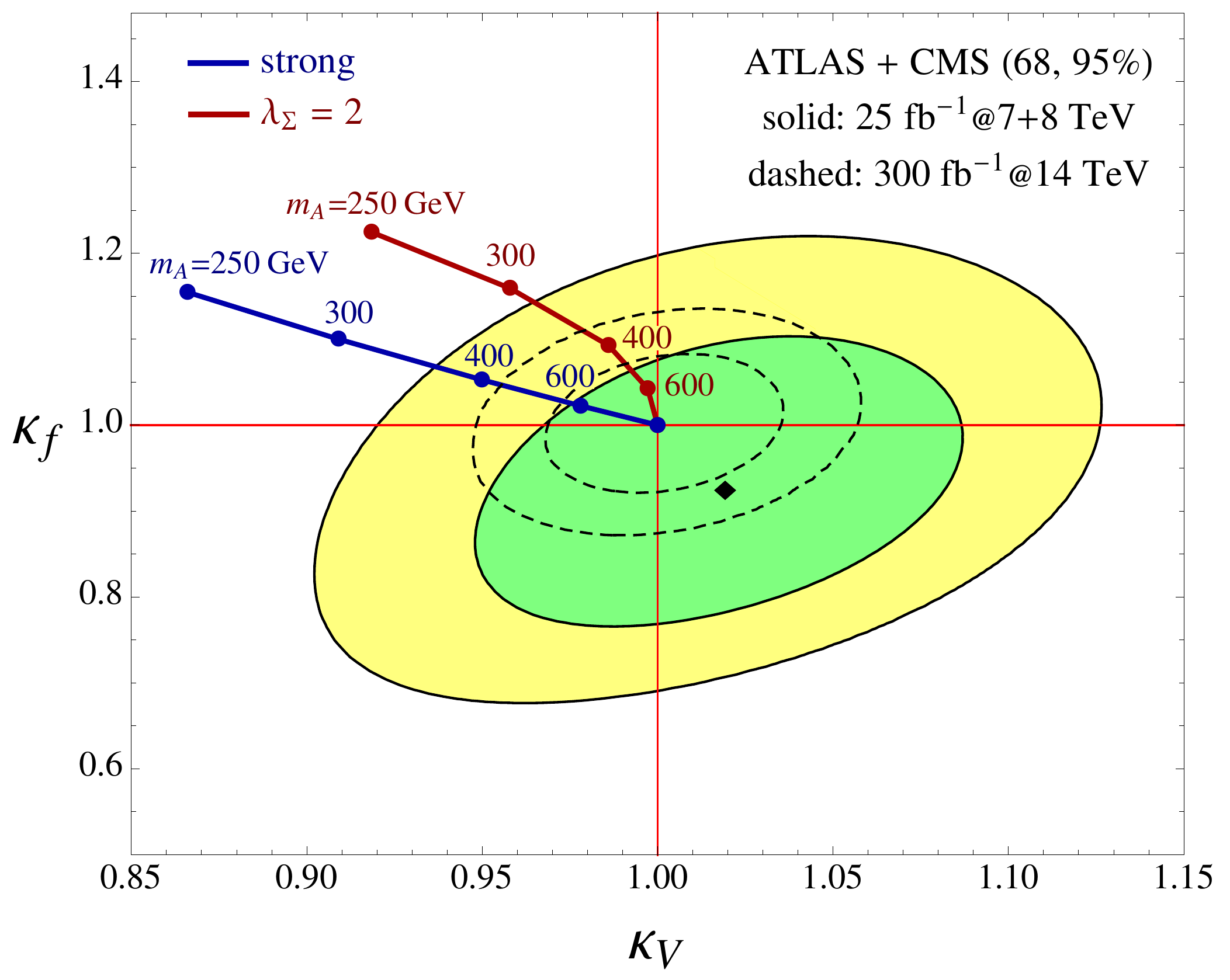} 
\caption{\small{Higgs couplings from ATLAS and CMS, with model trajectories following varying values of the light CP-odd scalar mass; in each case we set the self-coupling of $H$ to zero in the potential, corresponding to $\tan \beta = 1$, and take $\la_\Si = 2$ in the perturbative case.  We show the present status at 68 and 95\% CL, with best fit indicated by a diamond, along with projections for measurements at the 14 TeV LHC assuming injection of a SM Higgs signal.}}
\label{fig:VF}
\end{figure}

We first consider the direct constraints on the $A^0$ and $H^+$ particles of the strongly-coupled model.  There are indirect constraints from $b\to s \gamma$ and 
the combined Higgs coupling fit using current results from ATLAS and CMS.
The Higgs coupling fit requires $f< 72\GeV$.
The $b\to s\gamma$ limit is both more model dependent and weaker than the coupling fit, so 
we do not present it in the following plots.  
The relevant direct searches are $A^0\to Zh, A^0\to \tau\tau, t\to H^+b\to (\tau^+ \bar{\nu}) b$,
which are detailed in the appendix.  
The constraints on the parameter space are illustrated in 
Fig.~\ref{fig:strongpseudolimits} where we compare $\la_H$ to the standard model value 
for a 125 GeV Higgs, $\lambda_{\text{SM}}$.  
The charged Higgs search rules out the range below $m_{H^\pm} = 160\GeV$, where the analysis stops due to the limited phase space in the top decay.  The $\tau\tau$ search is then the strongest direct search up to about 220 GeV, where the analysis loses sensitivity.  
For the range $225$-$460$ GeV, the $A^0\to Z h$ search constrains most values of $\la_H$.
The Higgs coupling measurements complement the direct searches, 
by improving the constraints in the region where $A^0\to \tau\tau$ is
the most sensitive direct search. The Higgs coupling constraints depend only on $m_A$ in the limit $\la_H = 0$,
as shown in Fig.~\ref{fig:VF}.

To interpret the constraints in terms of induced EWSB, we note that for
$\la_H \lsim 0.7 \la_\text{SM}$, the Higgs mass-squared parameter is positive.
This is therefore the region where EWSB is induced by a tadpole. 
The viable parameter space for induced EWSB thus requires $m_A \gsim 460$ GeV, 
while for lower $m_A$ masses, the constraints require the tadpole to be supplemented by
a negative mass-squared for the Higgs doublet.

We also made projections for the sensitivity for the 14~TeV run of the LHC,
details of which are given in the appendix.
The search for $A^0 \to Zh$ is so sensitive that it can nearly probe the entire allowed parameter
space with only 20~fb$^{-1}$, as shown in the blue dashed line in 
Fig.~\ref{fig:strongpseudolimits}.
(We cannot project the $A^0 \to Zh$ search below $m_A = 225\GeV$, the smallest mass considered 
in the current experimental analysis, but it is clear that the search has sensitivity down to $m_A\gtrsim m_h + m_Z$.)
This will therefore be an early discovery mode at the 14~TeV LHC in this class of models.  We also projected how sensitive the direct searches for $\tau\tau$ and $t\bar{t}$ will be with 
300~fb$^{-1}$ at the 14 TeV LHC.  Note that  the $A^0 \to t\bar{t}$ search is still not sensitive and so is not included in the plot.

The Higgs coupling fit improves only marginally for 300 fb$^{-1}$ \cite{Dawson:2013bba}.
Assuming that the central value is equal to the standard model, we find a constraint
of $f<59$~GeV.
The reason for this rather weak improvement can be seen in Fig.~\ref{fig:VF}. 
The current best fit point shows a mild preference for a reduced fermion coupling and an 
enhanced vector coupling compared to the SM, which is the opposite of what the model predicts
(see Eq.~\eqref{eq:hcoupstrong}).
That is, the current bound is stronger than the expected limit, 
and the projected bounds for the 14~TeV LHC are weaker than would be inferred from a 
naive rescaling of current exclusions.
This is reflected by the relatively weak projected bound in dashed gray in 
Fig.~\ref{fig:strongpseudolimits}.

In the strongly-coupled model, we expect additional effects from the
production of technihadron states.  
We consider vector resonances (``technirhos'')
as an example, motivated by the fact that these
are prominent on the phenomenology of QCD-like theories.
The largest production of technirhos at the LHC is generally Drell-Yan production of the charged
$\rho$, which arises from mixing between the $\rho^+$ and the $W$.
The mixing term is proportional to $g / 2g_\rho$, so the production rate is suppressed
for large $g_\rho$ due both to the increased $m_\rho$ and decreased coupling strength.
The vector resonances will decay preferentially to the (mostly) composite
pseudoscalars.
The decay $\rho^+ \to H^+ A^0$ will therefore dominate if kinematically open,
but the constraints on the pseudoscalars generally force them to be
sufficiently heavy that this mode is unlikely to be open.
This leaves the decays $\rho^+ \to W^+ A^0$ or $Z H^+$ 
and $\rho^+ \to W^+ Z$.

\begin{figure}[t!]
\centering
   \includegraphics[width=.48\textwidth]{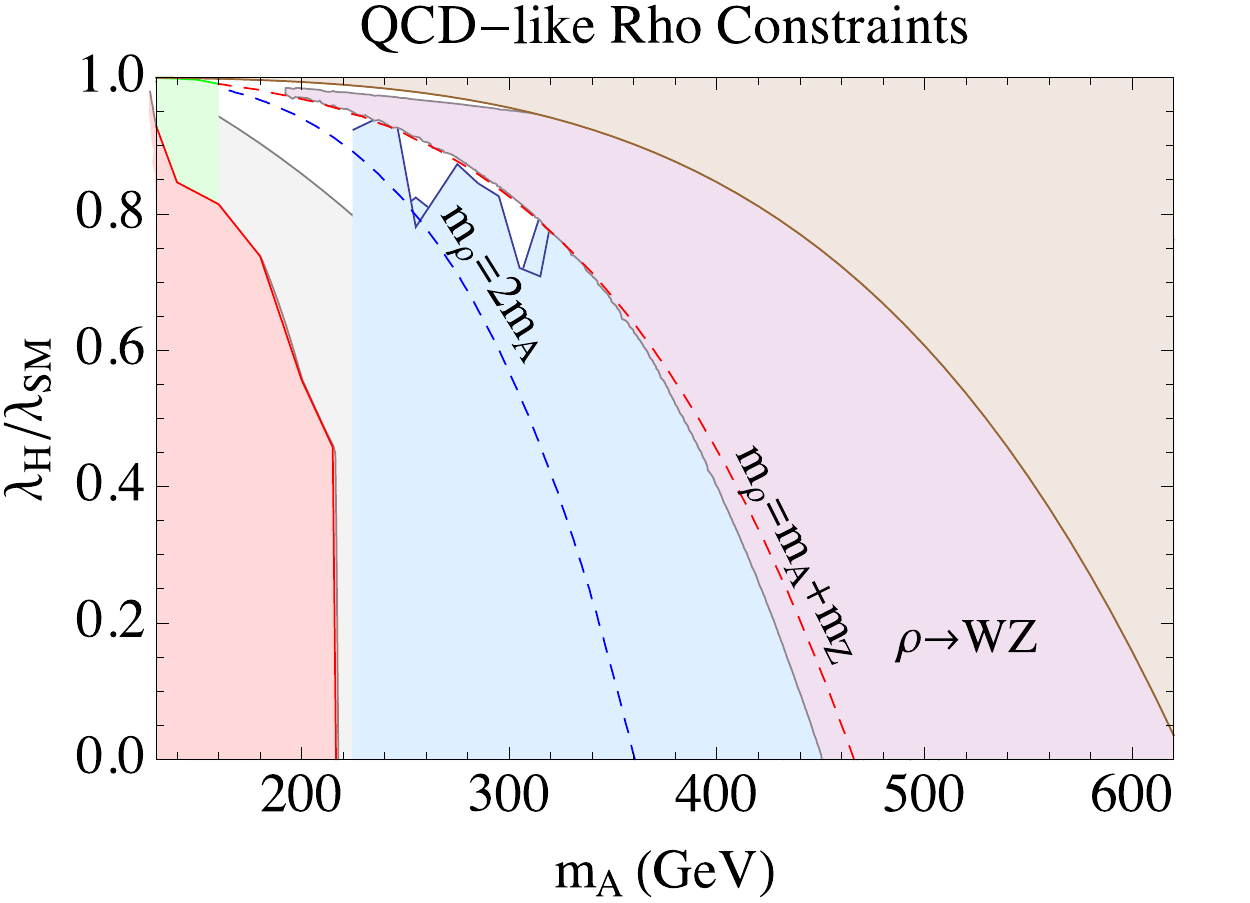} \quad 
   \includegraphics[width=.48\textwidth]{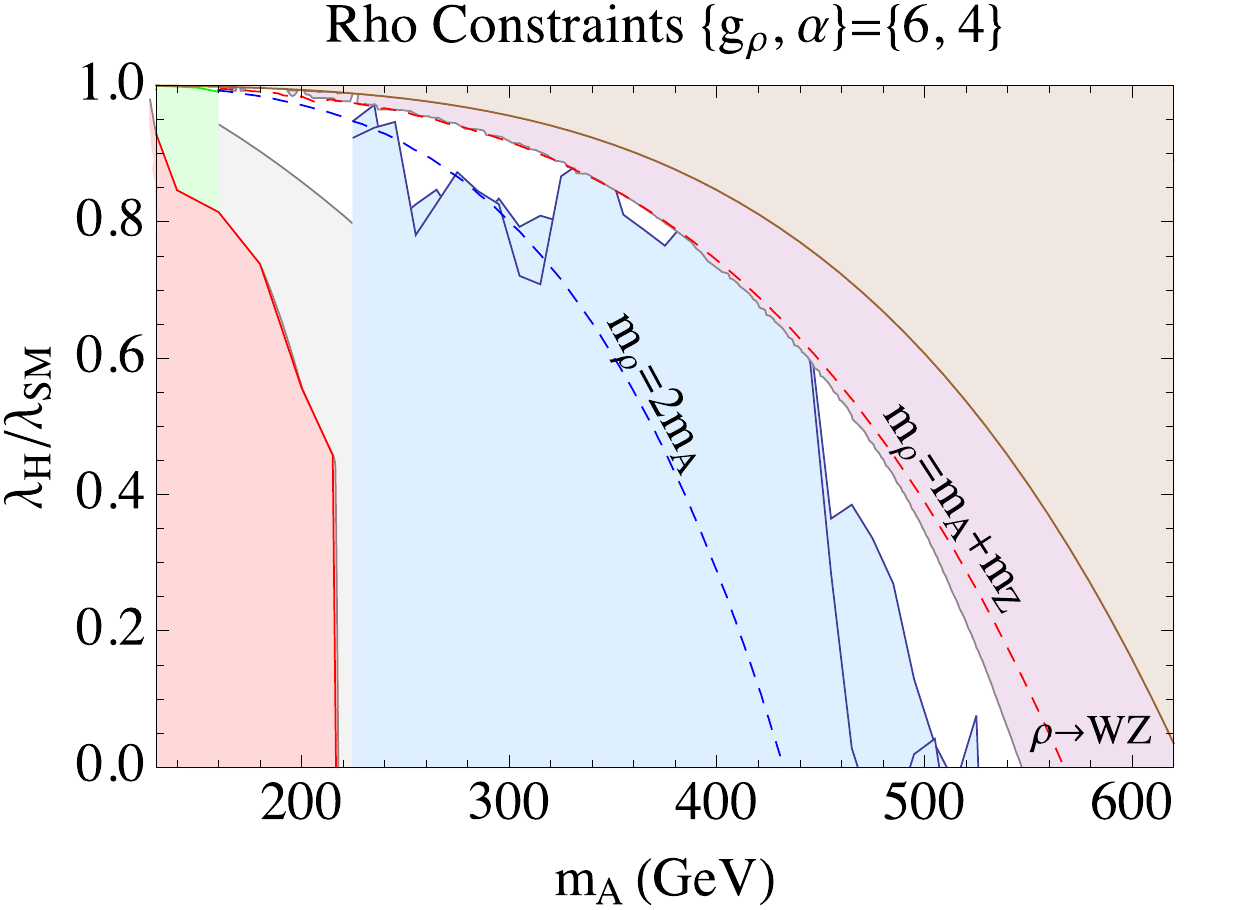}
      \includegraphics[width=.48\textwidth]{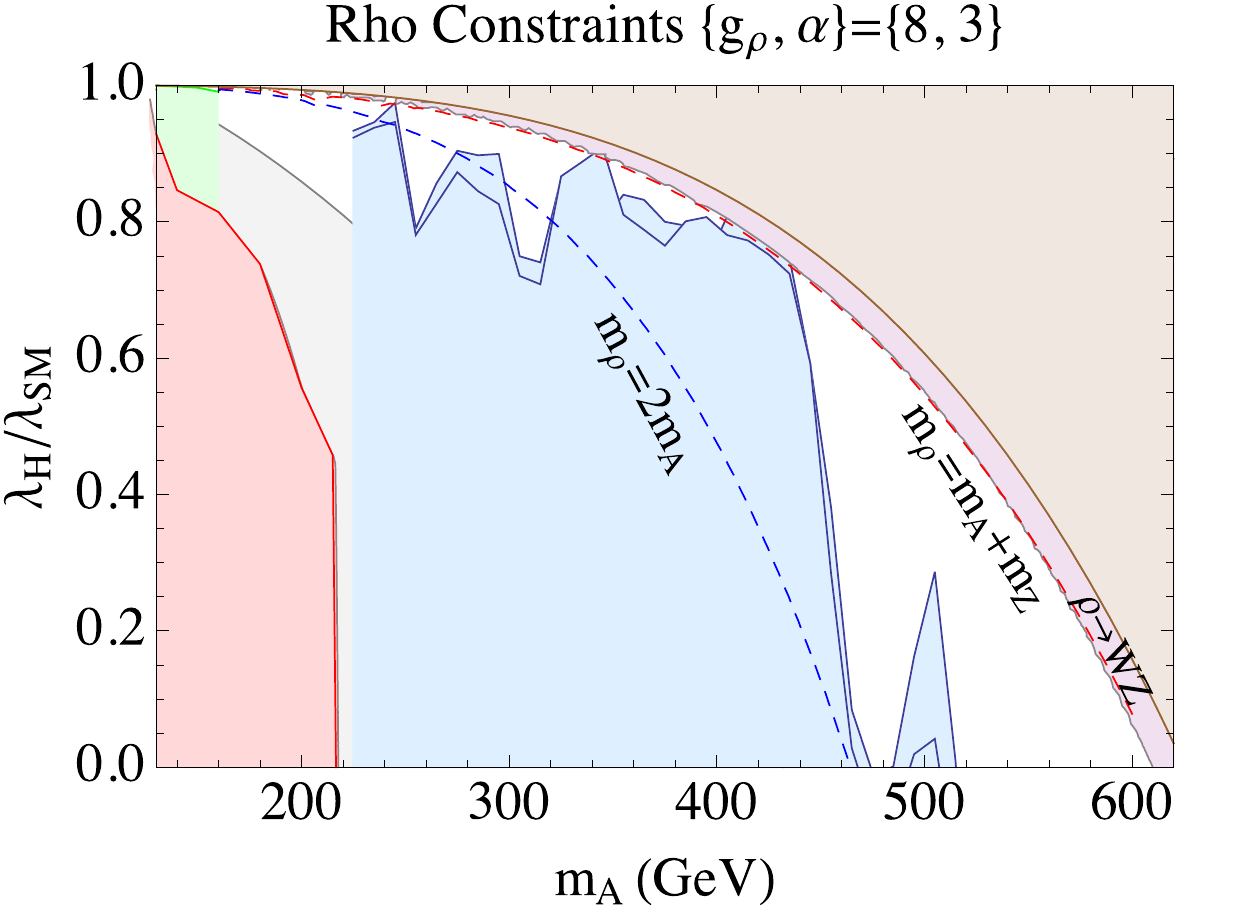}
\caption{\small{Additional constraints due to technirho production.  For unlabeled contours, see Fig.~\ref{fig:strongpseudolimits} for labeling.  The new constraint is the multilepton search for $\rho^+\to W^+Z$.  The kinematic thresholds for $\rho^+ \to H^+ Z, H^+ A^0$ are shown in dotted lines, where the  decay is open to the left of the line. We also include the increased production of $A^0$ from rho decays in the constraints for $A^0\to Zh$ and $A^0\to \tau\tau$, as illustrated by the additional parameter space excluded by those searches.}}
\label{fig:rholimits}
\end{figure}

As an illustration of some of the additional constraints from the technirho, 
we consider the benchmarks of a QCD-like rho $(g_\rho,\alpha)=(6.4,1.7)$ and two more 
strongly-coupled scenarios $(g_\rho,\alpha)=(6,4)$ and $(g_\rho,\alpha)=(8,3)$.
The constraints are shown in Fig.~\ref{fig:rholimits}.  
Here, we have added the CMS multilepton search for $\rho\to WZ$ to the parameter space plots, which constrains the magenta shaded region to the right.

\begin{figure}[t!]
\centering
   \includegraphics[width=.6\textwidth]{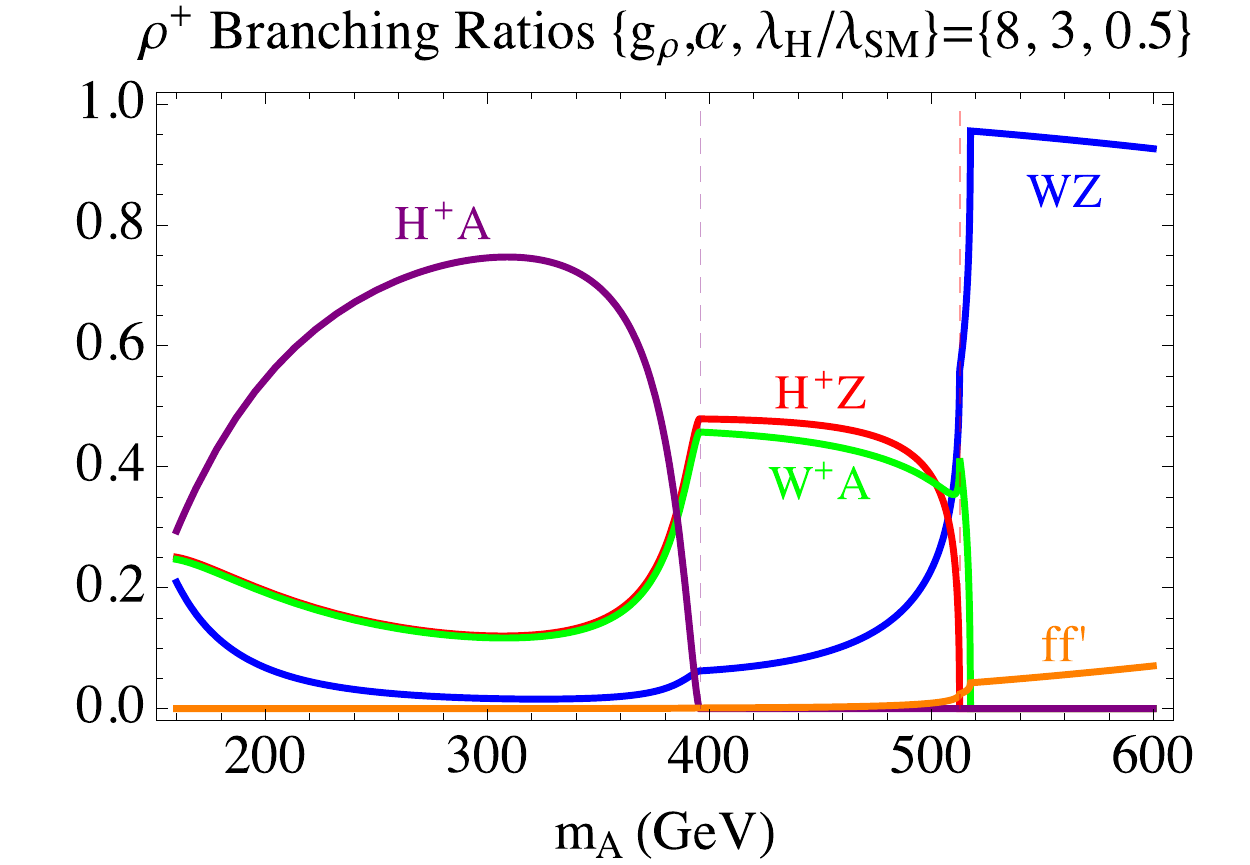}
\caption{\small{Branching ratios of the charged technirho for $g_\rho=8,\alpha=3,\la_H=\la_{\mathrm{SM}}/2$.  The mass of the technirho decreases as $m_A$ increases and thus these strongly interacting modes close for large $m_A$. To illustrate this behavior, the kinematic thresholds of $H^+A$ and $H^+Z$ are labeled as vertical dashed lines.}}
\label{fig:rhoBRs}
\end{figure}

The behavior of these constraints can be understood by looking at the technirho branching ratios, an example of which is shown in Fig.~\ref{fig:rhoBRs}.  As one goes to higher $m_A$, $f$ goes down, decreasing the $\rho$ mass.  Thus, at some point, for kinematic reasons,  the technirho can only decay into $WZ$ and SM fermions $f \bar{f}'.$  The $WZ$ search is quite strong and thus rules out this region.  We have also checked that $W'$ searches for decays $\ell \nu$ set weaker constraints than $WZ$.  On the other hand, as one goes to lower $m_A$, the $\rho$ mass increases, opening up decays to the pseudoscalars.  Once the decays are open, they tend to dominate due to the large $g_{\rho\pi\pi}$ coupling.  The kinematic thresholds where $H^+ Z, H^+ A^0$ open up are shown in dashed lines in Figs.~\ref{fig:rholimits}, \ref{fig:rhoBRs}, which explains the dropoff in sensitivity to $WZ$.  In Fig.~\ref{fig:rholimits} we also include the increased production of $A^0$ from technirho decays in the constraints for $A^0\to Zh, A^0\to \tau\tau$, as illustrated by the additional parameter space excluded by those searches.  These benchmarks give a flavor of the constraints.  For  a QCD-like rho the constraints are complementary to the Higgs coupling and $A\to Zh$ constraints which together almost completely exclude the full parameter space. However, the more strongly-coupled benchmarks show that increases in $g_\rho$ or $\alpha$ push the technirho heavier,  weakening the limits on parameter space, allowing a larger range where interesting technirho phenomenology of multistep cascades is allowed.

\begin{figure}[t!]
\centering
   \includegraphics[width=.48\textwidth]{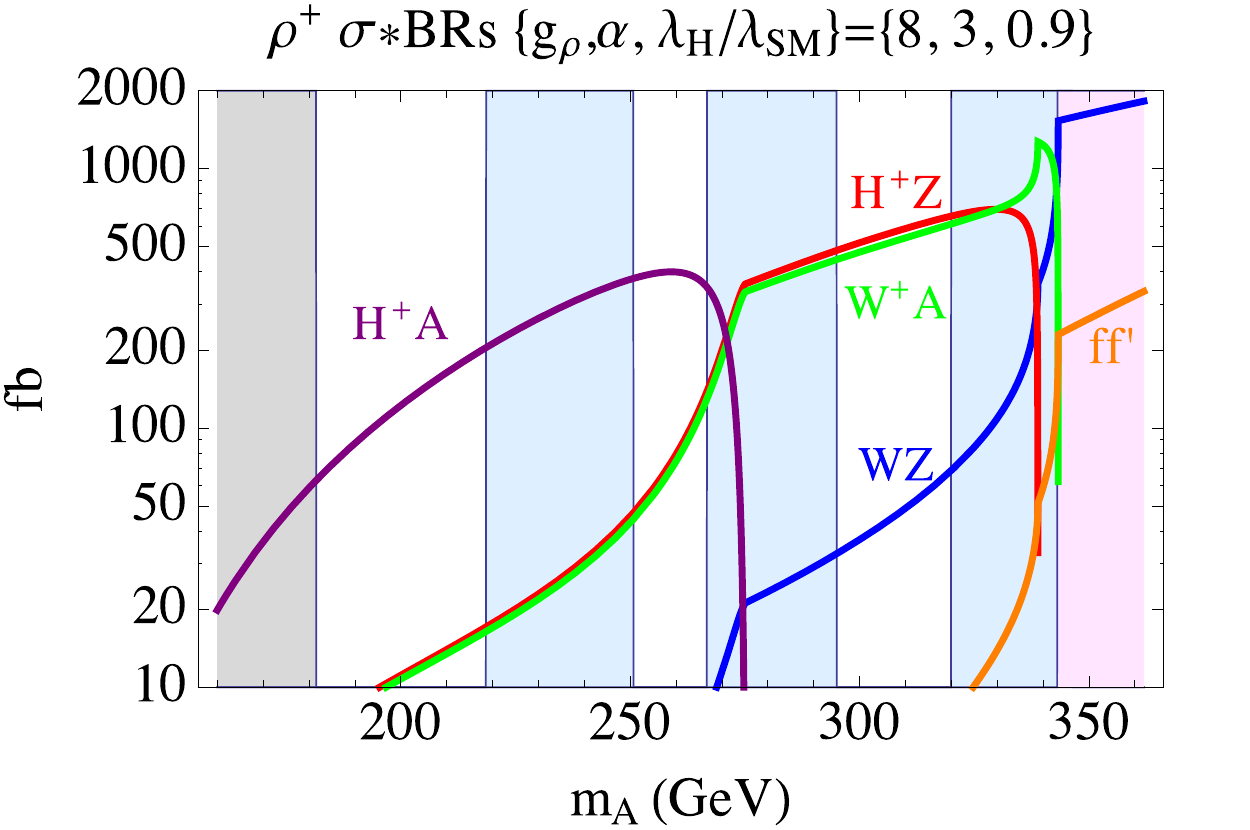}  \includegraphics[width=.48\textwidth]{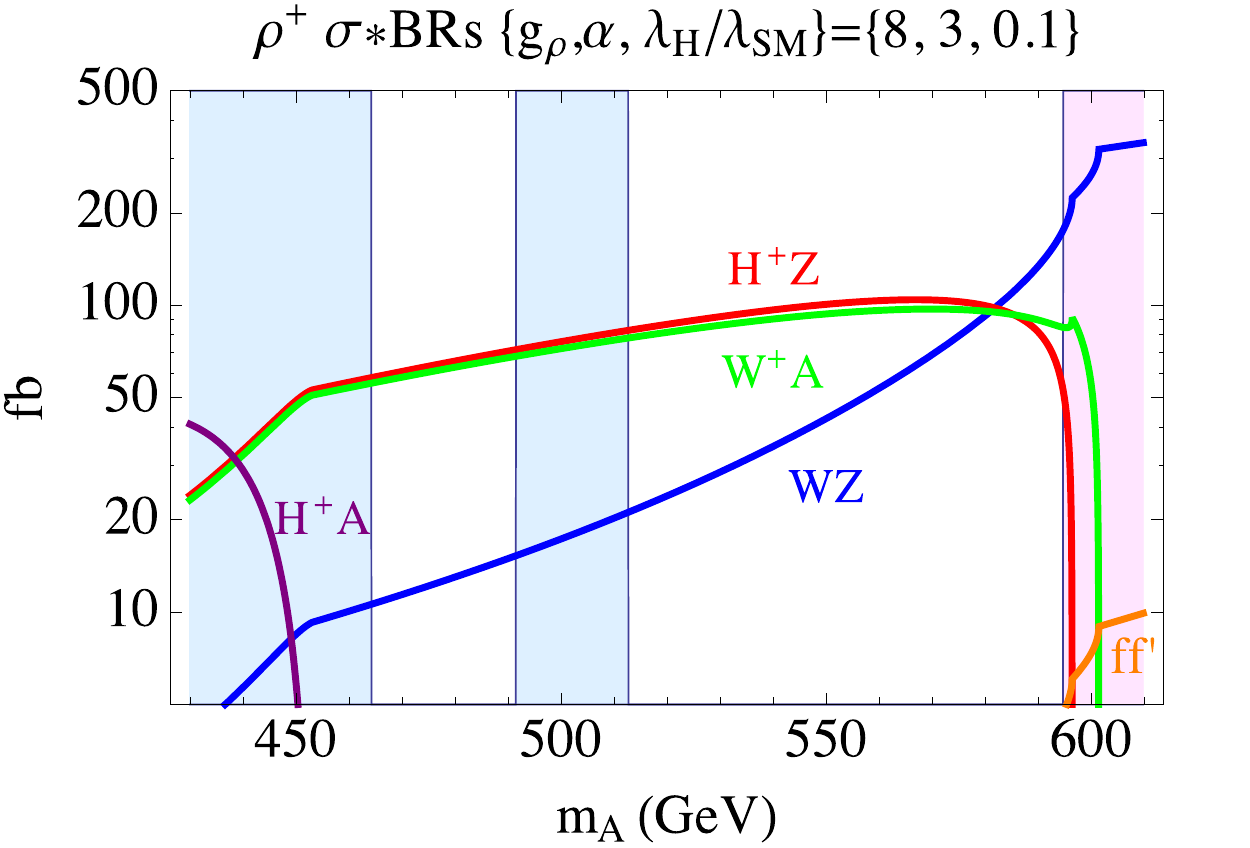}
\caption{\small{Cross section times branching ratios for the charged technirho at the 14 TeV LHC for $g_\rho=8,\alpha=3$, and $\la_H/\la_{\mathrm{SM}}=0.9 \text{ and } 0.1$. Exclusions from the $A\to Zh$, $\rho\to WZ$ and Higgs coupling fits are denoted by shaded regions with coloring similar to Fig.~\ref{fig:rholimits}.}}
\label{fig:rhoxsecBRs}
\end{figure}

Looking ahead to future searches, given that the allowed parameter space requires heavy masses,  the pseudoscalars will typically decay into
$H^+\to t\bar{b}$ and $A^0 \to Zh, \bar{t}t$.  
Hence, the mixed decays of the technirho end up as 
\begin{eqnarray}
\rho^+ &\to& W^+ A^0 \to W^+ (Z h) \text{ or } W^+ (t\bar{t}),\\
\rho^+ &\to& H^+ Z \to (t\bar{b}) Z.
\end{eqnarray}
Examples of the rates for these technirho cross sections are given in Fig.~\ref{fig:rhoxsecBRs}, which show that the mixed decays can have cross sections as high as 700 fb.
There are currently no dedicated searches for such cascades, although they can produce a
signal in multilepton searches.
The neutral resonances have smaller production cross sections
and also a simpler phenomenology.  
They couple strongly only to charged states and therefore
the mixed decays are
\beq
\rho^0 \to W^- H^+ \to W^- (t\bar{b}),
\eeq
plus the charge conjugate state.
For a much smaller part of allowed parameter space, it is possible for the technirho to decay into two pseudoscalars.  Here the decays are 
\begin{eqnarray}
\rho^+ &\to& H^+ A^0 \to (t\bar{b}) (Z h)\text{ or }  (t\bar{b}) (t\bar{t}),\\
\rho^0 &\to& H^+ H^- \to (t\bar{b})(\bar{t}b).
\end{eqnarray}
The cross sections for these decays into pseudoscalar pairs have typically smaller rates as can be seen in Fig.~\ref{fig:rhoxsecBRs}.
Direct technirho decays to final states involving the light Higgs are strongly suppressed by the small mixing between the $\rho$ triplet and the light gauge bosons. For example, for $\rho^+ \to W^+ h$ we find
\beq
\frac{\Gamma(\rho^+ \to W h)}{\Gamma(\rho^+ \to W^+ Z)} \sim \left(\frac{v^2}{f^2}\frac{g^2}{g_\rho^2}\right)^2 \sim 10^{-2}\,,
\eeq 
where we took $f^2/v^2\sim 0.1$ and $g/g_\rho \sim 0.1$ as rough estimates of the parameters.  These decay widths are included in the plot of Fig.~\ref{fig:rhoBRs}, which illustrates the rareness of such decays.

In the strongly-coupled scenario, we see that there are potential signals with multiple electroweak gauge bosons and heavy flavor quarks.  This occurs generically since the pseudoscalars have an upper bound on their mass which allows them to be kinematically accessible to technihadron decays.  At the same time, the small amount of EWSB in the technicolor sector suppresses the couplings for the pseudoscalars, allowing them to be consistent with direct searches, but still allowing for them to decay into standard model states.  This rich phenomenology gives a crucial handle on uncovering the mechanism of induced EWSB.
%

\subsection{Weakly-Coupled Induced Electroweak Symmetry Breaking}
The parameter space of the weakly coupled simplified model is mainly constrained by LHC data, with additional constraints coming from $b\to s\ga$ and, to a much lesser extent, from the measurement of $R_b$ at LEP/SLD. The size of $\tan\be$ affects significantly the Higgs cubic coupling, but only has minor effects on the constraints. Therefore in the following we focus on $\tan\be=1$, and we will comment about the case $\tan\be=\infty$ at the end. The details of each experimental bound and of the method used to derive the 14 TeV projections are described in the appendix, where all the corresponding references can also be found.

\begin{figure}
\centering
\vspace{-3mm}
   \includegraphics[width=.8\textwidth]{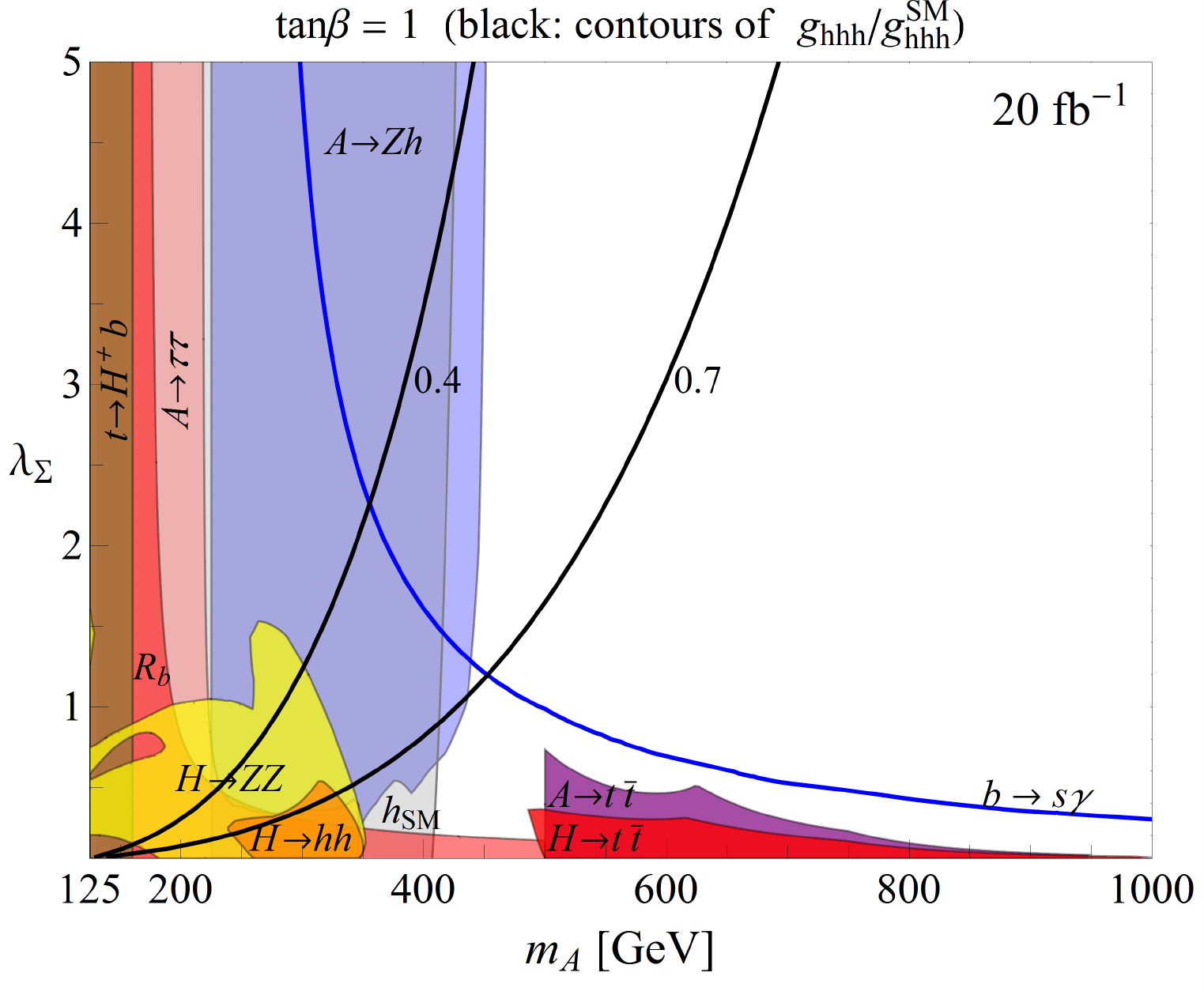} 
\vspace{-3mm}
   \includegraphics[width=.8\textwidth]{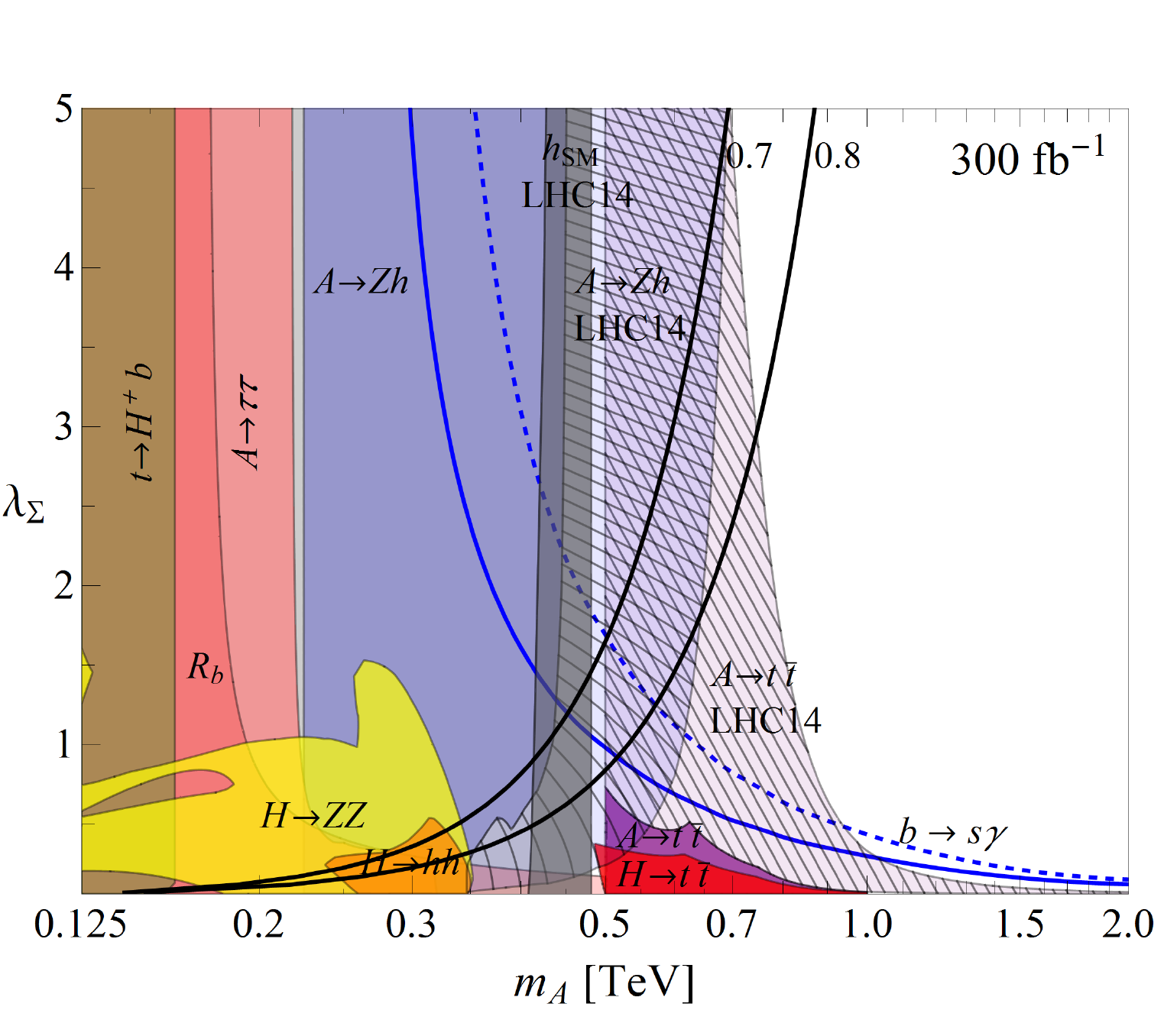}
\caption{\small{Current (top) and projected (bottom) constraints on the weakly coupled model with $\tan\be=1$. In the bottom figure, the hatching shows regions of parameter space that are presently open, but will be constrained by direct searches in $A^0\to Zh,\,A^0\to t\bar{t}$ at the 14 TeV LHC with $300$ fb$^{-1}$.} }
\label{fig:tanb1}
\end{figure}

\paragraph{$\boldsymbol {\tan  \beta = 1}$:}  
A summary of the current bounds for this case is shown in Fig.~\ref{fig:tanb1}(a). 
The strongest constraint comes from the search for $A^0\to Zh$, which excludes the mass range $225\;\mathrm{GeV} \lsim m_A \lsim 450\;\mathrm{GeV}$ for $\la_\Si \gsim 1$. For $m_A < 2m_t$ the decay $A^0\to Zh$ dominates, while above the $t\bar{t}$ threshold the branching ratio is small (see Fig.~\ref{fig:weakpseudobrs}) but the search has enough sensitivity to exclude masses up to $450\;\mathrm{GeV}$. 

The fit to the couplings of the light Higgs provides the second strongest constraint, giving 
$m_{A}\gsim 420$ GeV independently of $\la_{\Si}$.
This reflects the form of the couplings in Eq.~\eqref{hcouptb1}: the bound is driven by the 
$h\bar{f}f$ coupling, which is to good approximation independent of the auxiliary quartic coupling. 
The Higgs coupling constraints are shown in Fig.~\ref{fig:VF} for the representative value 
$\la_\Si=2$. As can be seen in Fig.~\ref{fig:tanb1}(b), the projection to $300$ fb$^{-1}$ of data at 14 TeV tightens the bound to $m_A \gsim 490$ GeV. Similarly to the strongly-coupled case, the projected bound is weaker than what would be naively expected by rescaling the current bound, because the current best fit point favors deviations from the SM in the directions opposite to those predicted by the model (see Eq.~\eqref{hcouptb1}), therefore the current bound is stronger than the expectation. To quantify the effect we also performed the 14 TeV projection by keeping the best fit points fixed to their current values, obtaining $m_A\gsim 550$ GeV.    

At present, the search for $A^0\to t\bar{t}$ only excludes a small portion of the parameter space 
with $m_A > 500$ GeV, but the 14~TeV projected sensitivity will cover a wide region and provide 
an additional important constraint to the model. 
Given the importance of this channel in directly testing induced EWSB, we urge the experimental collaborations to extend the search to lower resonance masses, ideally down to 
$m_A \gsim 2m_t$, where it would complement the sensitivity in the $A\to Zh$ search.
Finally, $A^0\to \tau\tau$ excludes the lower mass range $m_A \lsim 220$ GeV.  

The measurement of the $B\to X_{s}\ga$ branching ratio indirectly constrains the model, due to the 1-loop contribution of the charged Higgs. The bound is stronger at small $\la_\Si$, where the $\Si$ doublet is mostly responsible for EWSB and thus $f$ is large, which in turn enhances the coupling of the charged Higgs to fermions, see Eq.~\eqref{SCHpcoup}. In contrast, for larger values of the quartic the LHC bounds are stronger. Assuming the future measurement of the $B\to X_s\ga$ branching ratio to be limited only by the $\sim 5\%$ nonperturbative QCD uncertainty, we obtain a slightly stronger exclusion, shown as a blue dashed line in Fig.~\ref{fig:tanb1}. We emphasize that additional contributions to the loop amplitude, which were neglected here, could modify the bound, for example those from other SUSY particles.

Subleading constraints on the charged Higgs are obtained from the search for $t\to H^+ b$, which rules out $m_{H^+}< 160$ GeV, and from $R_b$. We also included for completeness the constraints on the heavy CP-even $H^0$, in the channels $ZZ,\,hh$ and $t\bar{t}$. All these bounds are relevant only at small $\la_{\Si}$, where the mass splitting between the triplet and the $H^0$ is moderate, and are subleading to the searches for the $A^0$ or to the Higgs couplings fit.  

\begin{figure}
\centering
\vspace{-3mm}
   \includegraphics[width=.8\textwidth]{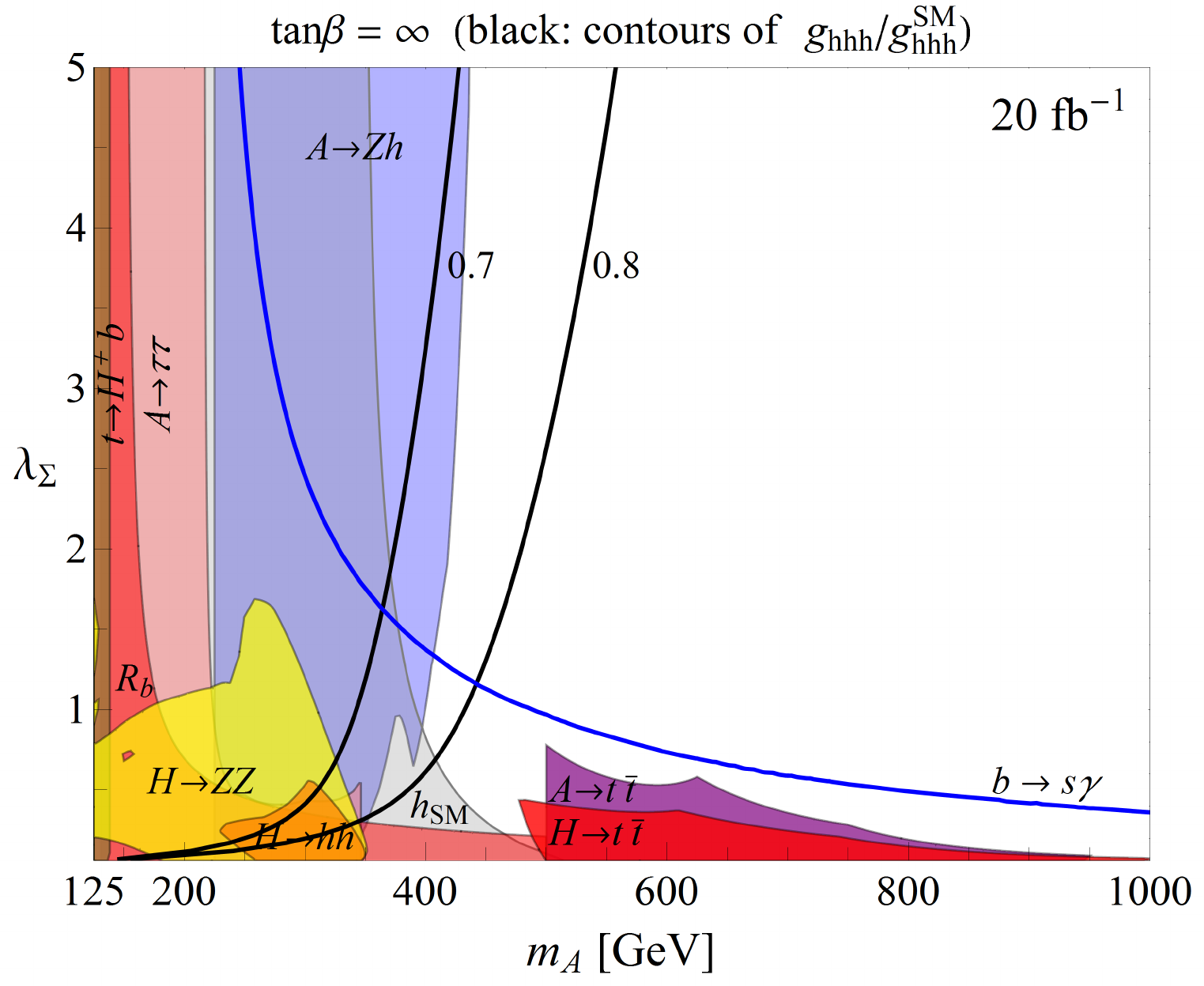} 
\vspace{-3mm}
   \includegraphics[width=.8\textwidth]{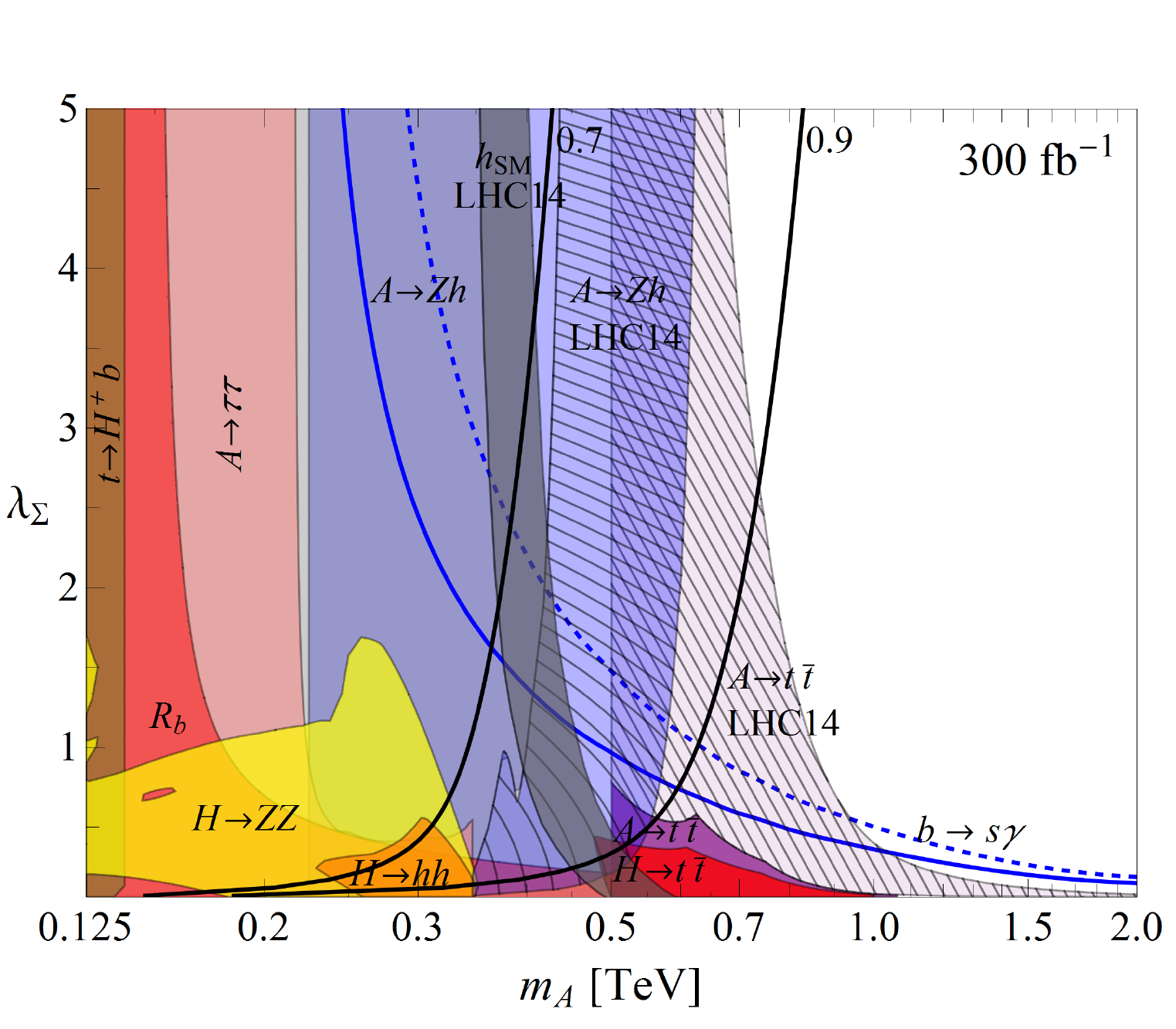}
\caption{\small{Current (top) and projected (bottom) constraints on the weakly coupled model with $\tan\be=\infty$. In the bottom figure, the hatching shows regions of parameter space that are presently open, but will be constrained by direct searches in $A^0\to Zh,\,A^0\to t\bar{t}$ at the 14 TeV LHC with $300$ fb$^{-1}$.}}
\label{fig:tanbinfty}
\end{figure}

In summary, the 8 TeV run of the LHC has constrained the parameter space of the weakly coupled model to $m_A \gsim 450$ GeV. Wide regions of parameters remain viable in which the EWSB is induced by a tadpole, as signaled by the suppressed cubic coupling, which can be as small as $40\%$ of the SM value and still be compatible with all current constraints. The 14 TeV run of the LHC will test further this idea, mainly via the direct search for signals of the light pseudotriplet, which is a peculiar feature of induced EWSB. The channels $A^0\to Zh,t\bar{t}$ have the capability to discover the neutral pseudoscalar in a wide mass range. Even if no signal is observed after $300$ fb$^{-1}$ of data, a deviation of $20\%$ in the Higgs cubic coupling will still be allowed. While such effect is challenging to measure at the LHC via double Higgs production, our results prove that it is in principle possible to observe a large deviation in the $h^3$ interaction consistently with the constraints on the other Higgs couplings and with direct searches. 
 
\paragraph{$\boldsymbol {\tan  \beta = \infty}$:}  
The constraints for this case are shown in Fig.~\ref{fig:tanbinfty}. 
The main difference compared to the case $\tan\be=1$ is the size of the Higgs cubic coupling, which is now larger because it receives a sizable contribution from the $D$-terms. On the contrary, the experimental constraints are qualitatively similar to those for $\tan\be=1$, albeit with some quantitative differences. First, the bound from the Higgs couplings fit has a nontrivial dependence on $\la_\Si$, which can be traced back to the form of the $h\bar{f}f$ coupling in Eq.~\eqref{hcouptbinfty}. Second, since the $D$-terms break the mass degeneracy $m_{H^{\pm}}=m_A$ by increasing the mass of the charged Higgs, the constraints from $t\to H^+ b, R_{b}$ and $b\to s\ga$ are slightly weaker. The 8 TeV $A^0\to Zh$ search has excluded up to $m_A \sim 430$ GeV, thus constraining the Higgs cubic coupling to be larger than $\sim 70\%$ of the SM value. Projecting to 14 TeV, if no signal is observed in the direct searches for $A^0\to Zh$ and $A^0\to t\bar{t}$ after $300$ fb$^{-1}$, then a cubic coupling as small as $g_{hhh}\sim 0.85 g_{hhh}^{\text{(SM)}}$ will be still allowed.

\section{Conclusions}
\label{sec:conclusions}

We have attempted to give a comprehensive survey of the constraints on the scenario
of induced EWSB,
in which the Higgs VEV is induced by a tadpole generated from an ``auxiliary'' Higgs sector.
Phenomenologically, this is a model where EWSB is nonlinearly realized at low energies,
while explaining why the observed Higgs boson is standard model-like.
The mechanism also gives an attractive possibility to generate a $125$ GeV Higgs in 
SUSY without fine-tuning.
We considered cases where the auxiliary Higgs sector is strongly-coupled as well
as perturbative.
Our main conclusions are as follows.

\begin{itemize}
\item Induced EWSB is consistent with all current bounds. 
The strongest constraints come from direct searches for the $A^0$ at the LHC.
$A^0\to Zh$ is highly constraining for $225< m_A < 450$ GeV and at higher masses, $A^0\to t\bar{t}$ constrains weakly coupled models. 

\item
The 14~TeV run of the LHC will have a wide discovery reach for this class of models.
In the strongly-coupled case, searches for $A^0 \to Zh$ can cover the entire allowed
range for this scenario with only 20~fb$^{-1}$.
For weakly-coupled models, there will still be parameter space open after 300~fb$^{-1}$.

\item
To obtain the full reach of LHC searches, it is important to extend them to cover the
full kinematic range.
In particular, the search for $A\to t\bar{t}$ for $m_A < 500\;\mathrm{GeV}$ can be a
discovery mode for the weakly coupled models.

\item
A significantly suppressed Higgs cubic coupling is compatible with  all other phenomenological
constraints.
In weakly-coupled models, we can have $g_{hhh} \sim 0.4 g_{hhh}^\text{(SM)}$ compatible
with current constraints, while with 300~fb$^{-1}$ of 14~TeV data we can still have
$g_{hhh} \sim 0.7 g_{hhh}^\text{(SM)}$.
For strongly-coupled models, currently there are no constraints on the smallness of $g_{hhh}$, however direct searches for the pseudoscalar at 14~TeV will already be sensitive to the region $g_{hhh} < 0.95 g_{hhh}^\text{(SM)}$.
 
\item 
In strongly-coupled models, there are additional potential signals from
vector resonances with masses $m_\rho \lsim 900\,\mathrm{GeV}$ decaying through
Higgs cascades, leading to 
final states involving electroweak gauge bosons, light Higgs and heavy SM fermions. 

\end{itemize}

Given the observation of a Higgs boson at 125 GeV with coupling close to the standard model value, it is natural to conclude that electroweak symmetry breaking is due to a single Higgs doublet.  In spite of this, we have shown that present constraints allow a much richer structure for the Higgs potential, where the Higgs VEV can be induced by additional sources of electroweak symmetry breaking.  Given our projections, we find that  next run of the LHC has significant reach in the parameter space of such models and thus still has much more to say on the mechanism of electroweak symmetry breaking.           

\section*{Acknowledgments}
We thank H.~Haber, A.~Kagan, A.~Martin, and M.~Peskin for discussions. S.C.~was supported in part by the Department of Energy under grant DE-SC0009945. J.G. was supported by the James Arthur Postdoctoral Fellowship at NYU. M.L., E.S. and Y.T. were supported by the Department of Energy under grant DE-FG02-91ER40674. J.G. and Y.T. thank the Aspen Center for Physics (National Science Foundation Grant No. PHYS-1066293) for hospitality during the completion of this work. E.S. thanks the Theory Group of CERN and the ITP of the University of Heidelberg for hospitality at various stages of this work.
\appendix{Appendix}
\label{sec:appendix}

\subsection{Direct searches at the 8 TeV LHC}
Below are listed the LHC searches that were used to set constraints on our models. Unless otherwise noted, they are based on a luminosity of $\sim 20$ fb$^{-1}$ at 8 TeV, combined in some cases with $\sim 5$ fb$^{-1}$ at 7 TeV.   

\paragraph{$\boldsymbol {t\to H^+b}$:} We use the CMS search for charged Higgs \cite{CMS-PAS-HIG-14-020}. The process considered is $t\bar{t}\to H^{+}b W b$ assuming the decay $H^{+}\to \tau \nu$, which gives a bound on $\mathrm{BR}(t\to H^{+}b)\times \mathrm{BR}(H^{+}\to \tau \nu)$ in the mass range $80\,\mathrm{GeV} < m_{H^{\pm}} < 160\,\mathrm{GeV}$. The same paper also reports on a search for charged Higgs with $m_{H^{\pm}}>180\,\mathrm{GeV}$ produced in association with a top quark and decaying to $\tau\nu$. However this search is irrelevant in our model, since $\mathrm{BR}(H^{+}\to \tau\nu)$ is very small for $m_{H^{\pm}}\gsim m_{b}+m_{t}\,$.

\paragraph{$\boldsymbol {A^0\to \tau\tau}$:} We use the ATLAS \cite{Aad:2014vgg} and CMS \cite{Khachatryan:2014wca} searches for scalars decaying to $\tau$ pairs. Both analyses quote a bound on $\sigma(gg\to A^0)\times\mathrm{BR}(A^0\to \tau\tau)$ in the mass range $90\,\mathrm{GeV} < m_{A} < 1\,\mathrm{TeV}$. For each mass point, we take the strongest between the CMS and ATLAS bounds.
\paragraph{$\boldsymbol {A^0\to Zh}$:} We use the CMS search for $A^0\to Zh \to \ell\ell b\bar{b}$ \cite{CMS-PAS-HIG-14-011}. We consider the bound on $\sigma(pp\to A^0)\times \mathrm{BR}(A^0\to Zh\to \ell\ell b\bar{b})$ shown in their Fig.~4 for the mass range $225\;\mathrm{GeV} < m_A < 600\;\mathrm{GeV}$. The dependence of $\mathrm{BR}(h\to b\bar{b})$ on the parameters of our model is taken into account. The CMS search for $A^0\to Zh$ in final states containing multileptons and photons \cite{CMS-PAS-HIG-13-025} and the ATLAS search for $Z+jj$ resonances \cite{ATLAS-CONF-2013-074} give weaker constraints.

\paragraph{$\boldsymbol {H^0\to hh}$:} For $m_{H^0}\lsim 380\; \mathrm{GeV}$ the strongest bound is given by the CMS $b\bar{b}\gamma\gamma$ search \cite{CMS-PAS-HIG-13-032}. We take the bound on $\sigma(pp\to H^0)\times \mathrm{BR}(H^0\to hh\to b\bar{b}\gamma\gamma)$ obtained from the `high purity' (2 or more $b$-tags) sample, reported in their Table~5 for the mass range $260 < m_{H^0} < 400\,\mathrm{GeV}$. The experimental bound is then compared to the same quantity computed in our model, taking into account also the modified BRs of the light Higgs. For $m_{H^0}\gsim 380\; \mathrm{GeV}$ the CMS search for resonances in the $hh\to b\bar{b}b\bar{b}$ final state \cite{CMS-PAS-HIG-14-013} has better sensitivity, but currently it does not exclude any region of the parameter space of the weakly coupled models.

\paragraph{$\boldsymbol {H^0\to ZZ}$:} We use the CMS search for Higgs bosons decaying to $ZZ\to 4\ell$ \cite{CMS-PAS-HIG-13-002}, where a bound on the cross section normalized to the SM one is given for the mass range $110\,\mathrm{GeV} < m_{H^0} < 1\,\mathrm{TeV}$. The bound from the $H^0\to WW$ channel \cite{Chatrchyan:2013iaa} is subleading and was not reported explicitly in our plots.
\paragraph{$\boldsymbol {A^0,H^0\to t\bar{t}}$:} We use the CMS search for resonances decaying to $t\bar{t}$ \cite{Chatrchyan:2013lca}. The results of the semileptonic resolved analysis, valid in the mass range $500\,\mathrm{GeV} < M < 1\,\mathrm{TeV}$ (with $M$ the resonance mass), are considered. The bound on the cross section quoted by CMS in their Fig.~2 refers to a spin-1 resonance, which has a smaller acceptance compared to a spin-0 particle because being $q\bar{q}$-produced, the vector is on average more boosted compared to the scalar, which is $gg$-produced. To take this effect into account, we computed the ratios of the acceptances of the CMS cuts for a CP-odd and -even scalar, divided by the acceptance for a spin-1 particle, and applied this correction to the bounds quoted by CMS for the $Z^{\prime}$. The acceptances were computed at parton level using the TopBSM MadGraph model, setting the couplings of each particle in such a way that the total width equals $10\%$ of the mass, corresponding to the experimental resolution on $m(t\bar{t})$ quoted by CMS. The couplings of the spin-1 to fermions were taken proportional to those of the $Z$. The acceptance ratio is for the pseudoscalar $\{1.2,1.3,1.3,1.2\}$ for $m = \{500,625,750,1000\} \,\mathrm{GeV}$ and for the scalar $\{1.6,1.4,1.3,1.2\}$ for $m = \{500,625,750,1000\} \,\mathrm{GeV}$.\footnote{We thank S.~Brochet and V.~Sordini for clarifications about the analysis.} 

\paragraph{$\boldsymbol {\rho^+\to W^+Z}$:} We use the ATLAS multilepton search for the technirho decay $\rho^+\to W^+Z$ \cite{Aad:2014pha}, imposing their listed limits on $\sigma(pp\to \rho^+)\times \text{BR}(\rho^+\to W^+Z)$ in the  range of $m_{\rho^+}$ from 200 to 1700 GeV.  Since the $\rho$ mass is reconstructed, we assume that other $WZ$ final states in a $\rho$ cascade do not fall into the same mass window.  

\paragraph{$\boldsymbol {\rho^+\to \ell^+\nu}$:} We use the CMS search for $W' \to \ell^+\nu$ \cite{Khachatryan:2014tva} using their combined limit on the leptonic decays $\sigma(pp\to W')\times \text{BR}(W'\to \ell^+\nu)$ for masses $m_{\rho^+}$ from 300 to 2000 GeV.  We chose to not use the ATLAS search \cite{ATLAS:2014wra} since it had worse limits at lighter  $W'$ masses.

\subsection{14 TeV projection}
Here we discuss the projection of the $8$ TeV $A^0\to Zh$ and $A^0\to t\bar{t}$ constraints to the $14$ TeV LHC. Since the experiments provide a bound on the cross section as a function of the assumed mass of the resonance, $\sigma^{8}_{S}(m_{A})$ (this includes the branching ratio into $Zh,t\bar{t}$), we obtain the projected $14$ TeV constraint as follows
\begin{equation}
\sigma^{14}_{S}(m_{A})=\sqrt{\frac{L_{8}}{L_{14}}}\,\sqrt{\frac{\sigma^{14}_{B}(m_{A})}{\sigma^{8}_{B}(m_{A})}}\,\sigma^{8}_{S}(m_{A})\,,
\end{equation}
where $L_{8,14}$ are the integrated luminosities at $8$ and $14$ TeV respectively, whereas $\sigma_{B}$ is the background cross section. We assume an integrated luminosity $L_{14}=300$ fb$^{-1}$. In the spirit of the Collider Reach tool \cite{colliderreach}, we assume that $\sigma_{B}$ simply scales with the parton luminosity of the main background process. In more detail:
\begin{itemize}
\item For $A^0\to t\bar{t}$ the main background is $pp\to t\bar{t}$, which is dominantly $gg$-initiated. Therefore $\sigma^{8}_{B}(m_{A})/\sigma^{14}_{B}(m_{A}) \sim \mathcal{L}_{gg}(m_{A}^{2},s_{14})/\mathcal{L}_{gg}(m_{A}^{2},s_{8})$\,, where 
$$\mathcal{L}_{ij}(M^2,s) = \tau \int_{\tau}^{1}\frac{dx}{x}f_{i}(x, M^{2})f_{j}(\tau/x, M^2)$$ 
is the parton luminosity ($\tau \equiv M^2/s$). For $\sigma_{S}^{8}$, in the relevant range $500\,\mathrm{GeV} < m_{A} < 1\,\mathrm{TeV}$ we take the expected $8$ TeV $Z^{\prime}$ limit, rescaled by the ratio of pseudoscalar/vector acceptances as described above. 
\item For $A^0\to Zh \,(\to \ell\ell b\bar{b})$ the main background is $pp\to Z+\mathrm{jets}$, which is mainly $q\bar{q}$-initiated. Therefore we take $\sigma^{14}_{B}(m_{A})/\sigma^{8}_{B}(m_{A}) \sim \mathcal{L}_{q\bar{q}}(m_{A}^{2},s_{14})/\mathcal{L}_{q\bar{q}}(m_{A}^{2},s_{8})$. For $\sigma_{S}^{8}$ we take the expected limit in the mass range covered by the CMS analysis, $225\;\mathrm{GeV} < m_{A} < 600\;\mathrm{GeV}$, whereas for $m_{A}>600\;\mathrm{GeV}$ we conservatively use the expected limit at $600\;\mathrm{GeV}$. 
\item For $A^0\to \tau\tau$, the main background depending on the channel is either $Z\to \tau\tau, \mu\mu$ (see \cite{Khachatryan:2014wca}) and thus we rescale the expected limit by using the luminosity ratio for $u\bar{u}$ to estimate the change in background.  
\end{itemize}
%

\subsection{Indirect bounds}
\paragraph{Light Higgs Couplings:}
The couplings of the light Higgs to other SM states are modified in all cases by a reduced vev, $\langle h \rangle < v$, and further in the weakly coupled models by the mixing between the CP-even neutral modes of $H$ and $\Si$.  Couplings to vectors and fermions have been measured at the LHC to a precision of order 10\% and 20\% respectively, providing indirect constraints on the enlarged scalar sectors of these models. 
For current constraints, we implement all Higgs production/decay channels reported by the ATLAS \cite{ATLAS-CONF-2014-009,Aad:2014eha} and CMS \cite{CMS:2014ega} collaborations in our model parameter spaces. For projections at 14 TeV, we adhere to expectations quoted in \cite{Dawson:2013bba} with uncertainties on the vector and fermion couplings  reduced to order 4\% and 8\%, respectively. Current and projected constraints in the space of $\ka_{f,V}$ were shown in Fig.~\ref{fig:VF}, together with the trajectories of the strongly and weakly coupled models as functions of $m_A$. The weakly coupled case allows for a lighter isotriplet. This stems from the fact that the light Higgs couplings in the perturbative model, for a given $m_A$,  are further modified with respect to the strong case by the presence of an additional CP-even mode with a mass determined by $\la_\Si$ (see Eq.~(\ref{eq:tb1H})).

\paragraph{$\boldsymbol {b\to s\gamma}$:} The charged Higgs contributes to the $C_{7,8}$ operators for $b\to s\gamma$. The model then is constrained by the $B\to X_s\gamma$ search. Following the standard convention for a type-I 2HDM, the couplings of the charged scalar to fermions are written as
\begin{equation}
(2\sqrt{2}G_F)^{1/2}\frac{f}{v_H}\,\sum_{i,j=1}^3\bar{u}_{i}(m_{u_i}V_{ij}P_L-m_{d_j}V_{ij}P_R)d_jH^{+}+\mathrm{h.c.}\,.
\end{equation}
Assuming the flavor structure in $V_{ij}$ to be aligned to the SM, $b\to s\gamma$ sets a direct constraint on $f/v_H$. Using the $95\%$ C.L. exclusion bound in the $(v_H/f,\,m_{H^{\pm}})$ space shown in Fig.~8 (right) of Ref.~\cite{Hermann:2012fc}, we derive the bound in the $(m_A\,,\la_{\Si})$ plane. Since Fig.~8 of Ref.~\cite{Hermann:2012fc} is limited to $m_{H^{\pm}} < 1$ TeV, we do a simple extrapolation of the exclusion bound to larger masses.

The improvement of the measurement of the $b\to s\gamma$ branching ratio is limited by the irreducible nonperturbative QCD uncertainty, which is believed to be $\simeq 5\%$.\footnote{We thank D.~Straub for discussions about this point.} Comparing to the current uncertainty $\simeq 10\%$, even assuming the future Belle-II measurement to have a negligible experimental error, the bound on the New Physics contribution can only be improved by a factor two. Since the amplitude of the one-loop diagrams is proportional to $(f/v_H)^2$, the future bound on $f/v_H$ is $\simeq 2^{1/4}$ times more stringent than the current constraint. We can then rescale the current bound to get an optimistic $b\to s \gamma$ projection.

\paragraph{$\boldsymbol {R_{b}}$:} The precise measurement at LEP and SLD of the quantity $R_{b}=\Gamma(Z\to b\bar{b})/\Gamma(Z\to \mathrm{hadrons})$ places an indirect bound on the model, since the charged Higgs contributes to $R_{b}$ at one loop. The theoretical prediction can be written as $R_{b}^{th}=R_{b}^{\mathrm{SM}}+\delta R_{b}$, where $R_{b}^{\mathrm{SM}}$ is the SM contribution including radiative corrections, and $\delta R_{b}$ is the new physics contribution, which depends on the parameters of the model and was taken from Ref.~\cite{Haber:1999zh}. The experimental value is $R_{b}^{exp}=0.21629\pm 0.00066$ \cite{Agashe:2014kda} and the SM prediction $R_{b}^{\mathrm{SM}}=0.21549$ \cite{Freitas:2012sy}.

\subsection{Theoretical predictions}
We summarize a few details about the theoretical predictions for production cross sections and branching ratios of $A^0,H^{\pm},\rho^\pm$ and $H^0$. Throughout the paper, the MSTW2008 PDFs \cite{Martin:2009iq} are used.
\begin{itemize}
\item To obtain the production cross section of the CP-odd $A^0$ at approximate NNLO, we multiply the exact $pp\to A^0$ cross section at LO in QCD times the NNLO $K$-factor computed for a CP-even Higgs (see below). We have checked that this procedure gives a result in agreement within $20\%$ with the results for 14 TeV $A^0$ production in Ref.~\cite{Harlander:2002vv}.
\item The production cross section of $\rho^\pm$ is computed at LO in QCD and multiplied times a constant factor $K=1.3$ that approximately accounts for higher order corrections.
\item The production cross section of the CP-even $H^0$ in gluon fusion is computed at NNLO in QCD using the code ggHiggs \cite{Ball:2013bra,Bonvini:2014jma}. The code gives the cross section for SM couplings, which we rescale to take into account the value of the $Ht\bar{t}$ coupling. For vector boson fusion we take the NNLO cross section for SM couplings \cite{Heinemeyer:2013tqa} and rescale it to take into account the value of the $HVV$ coupling.   

\item We include QCD corrections to the branching ratios of $A^0,H^{\pm},H^0$ into quarks, making use of the formulas given e.g. in Refs.~\cite{Djouadi:2005gi, Djouadi:2005gj}. Only tree-level, two-body decays are considered.
\end{itemize}

\bibliographystyle{utphys}
\bibliography{Induced_ref}
\end{document}